\documentclass[twocolumn]{article}

\usepackage[ruled,figure, linesnumbered]{algorithm2e}
\usepackage{subcaption}
\usepackage{graphicx}
\usepackage{amsmath}
\usepackage{listings}
\usepackage{hyperref}

\newcommand{\ie}{i.\@e.\@ }
\newcommand{\eg}{e.\@g.\@ }
\newcommand{\etal}{{\it et.\@al.\@}}

\newtheorem{theorem}{Theorem}[section]

\newtheorem{definition}[theorem]{Definition}

\newcommand{\defined}{\ensuremath{\doteq}}

\newcommand{\scalefig}{0.48}

\newcommand{\argmin}{\mathop{\mathrm{argmin}}}
\newcommand{\argmax}{\mathop{\mathrm{argmax}}}
\newcommand{\argmaxk}[1]{\mathop{\mathrm{argmax}[#1]}}
\newcommand{\argmink}[1]{\mathop{\mathrm{argmin}[#1]}}

\newcommand{\farness}{\ensuremath{F}}
\newcommand{\avgdist}{\ensuremath{A}}
\newcommand{\exptvalueapprox}{\ensuremath{E}}
\newcommand{\exptvalue}{\ensuremath{E^{*}}}
\newcommand{\optimalkmed}{\ensuremath{M^{*}}}
\newcommand{\approxkmed}{\ensuremath{M_{\mbox{\it method}}}}

\newcommand{\kmedian}{$k$-median}
\newcommand{\diameter}{\ensuremath{D}}

\newcommand{\degreeplus}{\ensuremath{\mbox{deg}^+}}

\newcommand{\degreef}{\ensuremath{\mbox{deg}}}
\newcommand{\kcore}{\ensuremath{\mbox{core}}}
\newcommand{\kcoreplus}{\ensuremath{\mbox{core}^{+}}}
\newcommand{\pagerank}{\ensuremath{\mbox{PageRank}}}
\newcommand{\voterank}{\ensuremath{\mbox{VoterRank}}}
\newcommand{\PRdamping}{\ensuremath{\delta}}

\newcommand{\mdegree}{{\tt degree}}
\newcommand{\mdegreeplus}{{\tt degree+}}
\newcommand{\mvrank}{{\tt VRank}}
\newcommand{\mprank}{{\tt PRank}}
\newcommand{\mcore}{{\tt core}}
\newcommand{\mcoreplus}{{\tt core+}}
\newcommand{\mhindex}{{\tt H-index}}
\newcommand{\mrandom}{{\tt random}}

\newcommand{\avgdegree}{\ensuremath{\langle d \rangle}}

\newcommand{\nn}[1]{{\em{#1}}}

\newcommand{\netscience}{\nn{ca-netscience}}

\title{\textsc{ Approximation algorithms  for k-median problems on complex networks: theory and practice}}
 \date{}
\author{Roldan Pozo \\National Institute of Standards \& Technology 
}
\begin{document}
 
 \maketitle

   \begin{abstract}
Finding the k-median in a network involves identifying a subset of k vertices that minimize the total distance to all other vertices in a graph. This problem has been extensively studied in computer science, graph theory, operations research, and numerous areas due to its significance in a wide range of applications.  While known to be computationally challenging (NP-hard) several approximation algorithms have been proposed, most with high-order polynomial-time complexity.  However, the graph topology of complex networks with heavy-tailed degree distributions present characteristics that can be exploited to yield custom-tailored algorithms.  
We compare eight algorithms specifically designed for complex networks and evaluate their performance based on accuracy and efficiency for problems of varying sizes and application areas. Rather than relying on a small number of problems, we conduct over 16,000 experiments 
covering a wide range of network sizes and k-median values. While individual results vary, a few methods provide consistently good results. We draw general conclusions about how algorithms perform in practice and provide general guidelines for solutions.

\end{abstract}

\medspace

\section{Introduction}

The \kmedian{} problem is an important and fundamental problem in graph theory, and various application areas. Given a connected network, it seeks to find $k$ vertices  which are the closest (in terms of average  distance) to the remaining vertices in the network graph. These $k$ vertices .  This is crucial in the spread of viral messages in social networks,  disease contagion in epidemiological models,  operation and distribution costs for goods and services, marketing and advertising,  design layout of communication networks, and other wide-ranging applications. 

Finding such an optimal set of vertices is referred to as the influence maximization problem~\cite{kempe2003maximizing} and numerous algorithms have been proposed to address this issue.  Although these algorithms were not explicitly designed for the $k$-median problem, they share mathematical similarities, as they attempt to find influential vertices that are similarly well-connected and can quickly disseminate information throughout the network. The level of  {\em influence} can be measured in various ways, typically employing diffusion models such as independent cascade model~\cite{granovetter1978threshold} or epidemiological models of the Susceptible-Infected (SI) and Susceptible-Infected-Recovered (SIR)~\cite{hethcote2000mathematics} types.  Section~\ref{sec:algos} illustrates how the $k$-median problem is a special case of these models.  A related problem, the $k$-center, minimizes the maximum, rather than the summation of distances to remaining nodes.  In practice, the $k$-medium is better adapted for distribution and throughput optimization as it minimizes the total distances traveled.

Formally, the $k$-median problem on network graphs has been shown to be NP-hard (in terms of $k$) by reduction to  the dominant cover problem~\cite{kariv1979algorithmic} and is computationally intractable for large network graphs.  The situation is difficult largely in part to the explosion of the solution space as the value of $k$ grows. A graph with $|V|$ vertices yields  $\binom{|V|}{k}$ possible $k$-element candidate sets.
Even a modest problem of $k=10$ for a graph with 1,000 vertices would yield $10^{23}$ vertex sets to examine.
In this paper, we consider real graphs containing {\em millions} of vertices with $k$ values up to 100, yielding solution spaces of over $10^{400}$ possibilities.  A brute-force approach, in such cases,  is clearly infeasible and approximation methods must be employed.

\section{Notation and definitions}

\begin{definition}
\label{def:kmedian}
{\bf k-median problem}: Given an integer $k$ and a connected graph $G = (V,E)$, find a set of $S$ of $k$ vertices that minimize the summation of distances  from $S$ to the remaining vertices of $G$. 
\end{definition}
Using  $d(v,u)$ as the distance (length of shortest path) between vertices $v$ and $u$, we can define the distance $d(v,S)$ between a single vertex and a vertex set $S$ as the minimum distance between $v$ and any vertex in $S$.  We can then define  the {\bf farness metric}  $F(S)$ as
\begin{equation}
\farness(S) \defined \sum_{v \: \in \: V-S} d(v, S)   \label{eqn:farness}
\end{equation}
 
Often, it is convenient to normalize this quantity by the number of external vertices and express this as an {\bf average distance}, $\avgdist(S)$ to the remainder of the graph as
\begin{equation}
\avgdist(S) \defined  \frac{\farness(S)}{ |V-S|} 
\label{eq:avgdist}
\end{equation}
as it is more intuitive and makes the comparisons of  these values more meaningful across graphs of varying sizes.  This is sometimes referred to as the {\bf normalized farness metric}. Note that a set $S$ that minimizes $\farness$ also minimizes $\avgdist$ and vice versa, as these are  scalar multiple of the other.  From here on we will utilize this  normalized farness metric, or simply the average-distance, $\avgdist(S)$ in the formulation of the original minimization problem, Def.~\ref{def:kmedian}, without loss of generality.

\medskip 
In this context, we can restate the {$k$-median} problem on a network graph $G$ to be the identification of a set of $k$-vertices which minimize the average distance to the remaining $|V|-k$ vertices: 
 \begin{equation}
\optimalkmed(k) \; \defined  \min_{|S|=k} \avgdist(S)
\label{eq:optimal}
 \end{equation}
 Thus,  $X = \{ v_1, v_2, \ldots v_k\}$  is a solution to the \kmedian{} problem if and only if
\begin{equation}
      \avgdist(X) = \optimalkmed(k)
\end{equation}
or, equivalently, if and only if
\begin{equation}
 X\in  \argmin_{|S|=k}  \avgdist(S)
\end{equation}
We refer to $\optimalkmed(k)$ as the true {\bf optimal value} of the \kmedian{} problem on a graph $G$, and let $M(k)$ denote an approximation to this solution using the methods described in Sec.~\ref{sec:algos}.  When necessary, we use  $\approxkmed(k)$ to avoid ambiguity.

The neighborhood of a vertex $a$  is the  set of its adjacent nodes and is given as $N(a) \defined  \{ b \in V \: | \: d(a,b) =1\}$.
The degree of a vertex, $\degreef(v)$, is given by $|N(v)|$. The $p$th level neighborhood shell is comprised of the vertices a distance $p$ away and is given as $N^{(p)}(a) \;\defined\; \{ b \in V \; | \;  d(a,b)  = p\}$. These definitions are extended to arbitrary subsets $A \subseteq V$ as  $d(A,b) \defined \min_{a \in A} d(a,b)$ and $N^{(p)}(A) \defined \{ b \in V \: | \: d(A,b) =p\}$.
If $S$ is a set, then $S|_k$ is the collection of all subsets of $S$ of size $k$.

Given these definitions, we can recast the farness metric  $F(S)$ in Eq.~\ref{eqn:farness} as a summation of neighborhood shells weighted by their distance:
\begin{equation}
                  F(S) \; = \; \sum_{p=0}^{\diameter} p \:  |N^{(p)}(S)|
\label{eqn:FSp}
\end{equation}
where $\diameter$ is the diameter of the graph.  
This alternate representation of $F(S)$ provides a connection to {\em influential nodes} network models described in Sec.~\ref{sec:algos}.

If $f$ is a function that maps domain $D$ to range $R$, $f: D \rightarrow R$, then it can be applied to any subset of $D$ to generate a set of values.  That is, for $S \subseteq D$,  $f(S) \defined \{ f(x), x \in S \}$.  For a finite set, $\max_{x \in S} f(x)$ returns the maximum value of $f$ over domain $S$, and
\begin{equation}
  \label{eq:argmax}
   \argmax_{x \in S} f(x)  \; \defined \; \{ s \in S,  f(s) = \max f(S)\}  
\end{equation}
denotes the elements of $S$ where the maxima of $f$ occur.  This notion can be extended to the top largest $k$ values of $f$, which can be considered an ordered list in decreasing order and represented as a sequence $T = ( x_1, x_2, \ldots x_k)$ where $f(x_1) \geq f(x_2) \geq  \ldots  \geq f(x_k)$.  As in the case of the conventional $\argmin$ function, this sequence may not be unique, due to ties. 
We define this extension of $\argmax$ as the set of all possible such sequences:
\begin{equation}
 \begin{aligned}
 \label{eq:argmaxk}
 \argmaxk{k}_{x \in S}  & f(x)  \;  \defined \;   \{ (x_1, x_2, \ldots x_k) : \\
                             &   f(x_i) \geq f(x_j)  \;\; \forall i < j \\
                             &    f(x_i) \geq f(s) \:\: \forall x \in  S \setminus \{x_1, \ldots x_{i-1}  \: \:  \}
\end{aligned}
\end{equation}
A similar definition can be described for $\argmin$ and $\argmink{k}$.  
While $\max f$ is a single value, $\argmax f$ is a set of values, and $\argmaxk{k} f$ is a set of sets of sequences.  Since any $x \in \argmax f$ is a maximum of $f$,  one can write  $x \leftarrow \argmax f$ to select one of its elements.  Similarly, $x \leftarrow \argmaxk{k} f$ selects a unique group of top-$k$ performers.

\section{Approximation algorithms and related problems}
\label{sec:algos}

The field of approximation methods to the \kmedian{} problem is quite large. Resse~\cite{allmethods} provides a comprehensive overview of over 100 methods from linear programming, genetic algorithms, simulated annealing, vertex substitution, and other approaches.  
 In general,  the algorithm with the best guaranteed approximation ratio is a local search and swap method~\cite{arya2001local} with provides a bound of $3 + 2/p$ where $p$ is the number of vertices simultaneously swapped. Its computation time is $O(n^p)$,  where $n$ is the number of vertices.  Thus, even for a quadratic-order complexity $O(n^2)$, which is quite limiting for large networks, the best guarantee we can get is a factor of four from optimal.
While these approaches were adequate for small networks, the  higher-order polynomial time complexity makes them unfeasible for networks with thousands or million of vertices~\cite{VoteRank}.

Instead, researchers have turned their attention to algorithms for finding effective spreaders in connected networks with heavy-tailed degree distributions, often employing an Susceptible-Infected (SI) or Susceptible-Infected-Recovered (SIR) model of spread~\cite{NewmanBook}~\cite{PorterBook}, where $I(t)$ denotes the number of infected nodes at time $t$, and $I(0)$ is the number of  initially infected nodes. The \kmedian{} can be thought of a special case of an  SI  model where the probability of an infected node transmitting the disease to a susceptible neighbor is 1.0, or an SIR model with the the probability of recovery for each node is 0.0.   In which case, the solution to the \kmedian problem can be thought of as maximizing the integral of the number of infected nodes over the propagation steps, starting with $k$ initial infected vertices.

That is, at each time step an infected node may transmit its disease to neighboring nodes, then starting with an infected set of vertices S and a probability of transmission of 1.0, the number of new infected nodes at each time step $t$, is simply the size of the neighborhood shell $N^{(t)}(S)$.  That is, 
\begin{equation}
  |N^{(t)}(S)|  \; = \;   I(t) - I(t-1)
\end{equation}
where we take each neighborhood shell to be a discrete time step.
In the formulation of epidemiological models, the continuous version of Eq.~\ref{eqn:FSp} can be expressed as
\begin{equation}
    \sum_{p=0}^{\diameter} t \:  |N^{(t)}(S)| \; \rightarrow \; \int_{0}^{\diameter} t \: I'(t) dt
 \label{eqn:FSIt}
\end{equation}
and integrating by parts, we have
\begin{align}
   \label{eqn:FSV1}
    \int_{0}^{\diameter} t \: I'(t) dt  \;  &= \;  t I(t) \arrowvert^{\diameter}_{0}  - \int_{0}^{\diameter} I(t) dt \\
    \label{eqn:FSV2}
                                                                      &= \;  |V| \cdot \diameter - \int_{0}^{\diameter} I(t) dt
\end{align}
Since $|V| \cdot \diameter$ is a fixed constant for each network, minimizing Eq.~\ref{eqn:FSV2} requires maximzing 
\begin{equation}
   \label{eqn:FSV3}
   \int_{0}^{\diameter} I(t) dt
\end{equation}
Hence, the set of maximal influencers $S$ that make $I(t)$  grown as quickly as possible, is also a solution to the k-median problem.  Thus, approximation methods for finding maximal influencers in networks can serve to find approximations to minimizing $F(S)$.

\medskip 

The algorithms described below represent methods used in the literature for solving these types of problems.  
Typically, they work by assigning each vertex a numerical value (centrality) which captures the affinity of this vertex to be a maximal influencer.  The vertices with the top $k$ values then become the candidate target set for the solution.
For each algorithm, we define $k$-median approximation as
\begin{equation}
X_{\mbox{\it method}}\:(k)  = (v_1, v_2, ... v_k) 
\end{equation}
and the corresponding  sum of distances as
\begin{equation}
 M_{\mbox{\it method}}^{*} \:(k) \; \defined \; A( \: X_{\mbox{\it method}}\:(k) \:  )
\end{equation}
with the intent that the relative error to the true solution is relatively small:
\begin{equation}
    \frac{M_{\mbox{\it method}}\:(k)  - M^{*}(k)}{ M^{*}(k)} \; \ll \; 1
\end{equation}

 \subsection{Degree ordering}

Approximating the \kmedian{} solution as the top $k$ hubs of the network is perhaps the most straightforward approach:
\begin{equation}
 X_{\mbox{\tt degree}}\:(k)  \;  \leftarrow \; \argmaxk{k} _{v  \in V} \degreef(v)
\end{equation}

The idea here is that the hubs (high-degree vertices) serve as efficient spreaders since they are connected to large number of neighbors.  This is countered by the notion that there may be significant overlap among their aggregate neighborhoods, i.e.  $|N(a_i)| + |N(a_2)| +  \ldots + |N(a_k)|$ can be significantly smaller than $|N(\{a_1, \ldots, a_k\})|$ with other vertices potentially covering the graph more effectively.

This is a common criticism of degree ordering for this problem, but our experimental results show that this may not be as critical an issue in practice (see Sec.~\ref{sec:conclusion}).

\subsection{Extended degree ordering}

A more sophisticated approach is the extended degree ordering, which measures the sum of degrees for neighboring vertices:
\begin{equation}
  \degreeplus(v) \: \defined \:  \sum_{x \in N(v)}  \degreef(x)
\end{equation}
and uses the top $k$ values as an approximation of the \kmedian{} solution:
\begin{equation}
 X_{\mbox{\tt degree+}}\:(k)   \; \leftarrow \; \argmaxk{k}_{v \in V}  \; \degreeplus(v)
\end{equation}
This is a semi-local algorithm, utilizing more information about the network's topology by analyzing the second-level neighborhood, i.e. neighbors of neighbors.  The motivation for this centrality measure is that it uses more information about the graph topology and can lead to an improved metric for identify candidate vertices for the set $S$.

\subsection{PageRank ordering}

PageRank is a variant of the eigenvalue centrality, which treats the network as a flow graph and values vertices with high eigenvalues.  It is the basis for some commercial web search engines.~\cite{NewmanBook}. Given a damping factor, $ 0 \leq \PRdamping \leq 1$, the PageRank centrality is given as the convergence of the iteration
\begin{equation}
  \pagerank(v) \: =  \:   (1 - \PRdamping) \;+ \; \PRdamping \sum_{u \in N(v)} \frac{ \pagerank(u)}{\degreef(u)}
\end{equation}
typically a value of $\PRdamping = 0.85$ is used in these calculations~\cite{NewmanBook}. %
The corresponding approximation for the \kmedian{} solution is
\begin{equation}
 X_{\mbox{\mprank{}}}\:(k)  \; \leftarrow \; \argmaxk{k}_{v \in V}  \; \pagerank(v)
\end{equation}

\subsection{VoteRank ordering}

A method developed by Zhang \etal~\cite{VoteRank}, is the VoteRank algorithm,  which uses an iterative voting methodology to determine the best influencer nodes.  Each vertex $i$ has a pair of values $(S_i, T_i)$ denoting the collective (incoming) votes from neighbors $S_i$ and the number of (outgoing) votes to give out in each voting round, $T_i$. At each voting round (complete pass through the graph) a vertex with the maximum (incoming) vote score is selected $(i^*)$  and its $(S_i^*, T_i^*)$ values are set to zero, effectively taking it out of future voting in subsequent rounds.  The neighbors of vertex $i^{*}$ have their respective $T_i$ votes reduced by a fixed value $f$, and the process is repeated until $k$ vertices are found.  In their paper, the authors use $f = 1/\langle d \rangle$, where $\langle d \rangle$ is the average degree of the graph, and this value is fixed throughout the algorithm.  Typically, one would choose $f$ such that $k f \ll 1$ but this implementation does not allow the $T_i$ values to go negative.  An outline of the algorithm is presented in Fig.~\ref{algo:VoteRank}.
The $k$-median approximation is given by
\begin{equation}
X_{\mbox{\mvrank{}}}\:(k) \; \leftarrow \; \argmaxk{k}_{v \in V}  \; \voterank(v)
\end{equation}
and we refer to this algorithm as \mvrank{}  in the comparisons.

\begin{algorithm}
  \BlankLine
 \KwData{Graph {\em G=(V,E), $f$, $N$}}
 \BlankLine
 \KwResult{ $R = (\voterank(v_1),  \ldots \voterank(v_N))$ }
 \BlankLine
  \BlankLine
   $S_v =\epsilon, \; T_v=1.0, \;\;  \forall v \in V$ \;
   
    \For{$i=1$ \KwTo $N$}
    {  
      \For{$v \in V$}
      {
         $S_v = \sum_{u \in N(v)} T_u$\;
      }
     $v^{*} = \argmax_{v \in V} S_v$ \;
     $R_i = v^{*}$ \;
     $S_{v^{*}}= 0.0 \; ; \; T_{v^*} = 0.0$ \;
     \For {$v \in N(v^*) $}
     {
        $T_v = \max(0.0, \; T_v - f)$\;
      }
    }
      
 \caption{outline of VoteRank algorithm}
\label{algo:VoteRank}
\end{algorithm}

\subsection{ Coreness ordering}

Another vertex centrality measure that has been proposed for finding effective spreaders is based on the degeneracy of network graphs.   The $i$-core of a graph is collection of connected components that remain after all vertices with degrees less than $n$ have been removed.  (This is often referred to as the $k$-core of a graph in the literature, but we use $i$ to avoid conflict with the $k$ used in the $k$-median formulation.) To compute the $i$-core of a graph, we remove all vertices of degree $i-1$ or less.  This process is repeated until there are no vertices of the graph with degrees less than $n$. The notion here is that vertices in higher value $i$-cores represent the inner backbone of the network, as opposed to lower-valued $i$-cores which lie at its periphery, and serve as better-connected vertices to efficiently spread information throughout the network.  The {\em core-number} of a vertex is the largest value $i$ which it belongs to the $i$-core of the graph.  

Although one could use this centrality to identify candidate vertices~\cite{kitsak2010identification}, one problem that has been noted is that the core values of the highest vertices are often the same and hence are not distinguishable to form a proper ordering~\cite{kcoreplus}.  To remedy this, a slightly extended centrality has been proposed that replaces the $i$-shell value of a vertex with the sum of its neighbors' core-number.   That is, if $c(v)$ is the core-number of $v$, then
\begin{equation}
C(v) \; \defined  \; \sum_{u \in  N(v)} c(u)
\end{equation}

\begin{equation}
X_{\mbox{\tt core}}\:(k) \; \leftarrow  \argmaxk{k}_{v \in V}  \; C(v) \; 
\end{equation}

We refer to this algorithm as \mcore{}.

\subsection{ Extended coreness ordering}

Extensions to the  $C(v)$ centrality have also been proposed~\cite{kcoreplus} as an improved measure for influence.  In a similar manner to $\degreeplus$,  {\it neighborhood coreness}, or $\kcoreplus$, uses the values of its neighbor's $\kcore$ centrality.
\begin{equation}
C^{+}(v) \; \defined  \; \sum_{u \in  N(v)} C(u)
\end{equation}

\begin{equation}
X_{\mbox{\tt core+}}\:(k) \; \leftarrow \;  \argmaxk{k}_{v \in V}  \; C^{+}(v) \; 
\end{equation}

We refer to this algorithm as \mcoreplus{}.

\subsection{ H-index ordering}

The Hirsch index or $H$-index~\cite{Hindex01}, originally intended to measure the impact of authors and journals by way of citations, has also been studied as centrality to measure  ranking of influence and its relation to other centralities~\cite{Hindex02}. The original measure for an author or journal was determined as the number of $n$ publications that have at least $n$ citations.
For example, an author with the following number of citations $[13,  16,  27, 9,  2]$ would have an $H$-index of 4. 
  In terms of a network graph, the Hirsch index of a vertex $v$, given as $H(v)$, can be represented as the maximal number of $n$ neighbors that each have at a degree of $n$ or more.  That is, if $h(v,n)$ is the number of neighbors of $v$ with degree at least $n$,
\begin{equation}
h(v,n) \defined \{ e \; |\;  \degreef(e) \geq n, \; e \in N(v) \} 
\end{equation}
then
\begin{equation}
H(v) \; \defined \;  \max_{n} \; \{  |h(v,n)| \geq n \}
\end{equation}
The $k$-median approximation can be then be given as
\begin{equation}
X_{\mbox{\tt H-index}}\:(k) \; \leftarrow  \argmaxk{k}_{v \in V}  \;  H(v) \; 
\end{equation}
This algorithm is referenced as \mhindex.

\subsection{Expected value (Random)}

The {\bf mean} average-distance of every $k$-element vertex set is simply the expected value of all possible combinations:

\begin{equation}
\exptvalue(k) \defined  \frac{1}{\binom{|V|}{k}} \: \sum_{|S|=k} \avgdist(S)
\label{eq:expvalue}
\end{equation}
That is, the average value of a {\bf random} guess chosen from a uniform distribution of all $\binom{|V|}{k}$ possible sets.
This can be computed exactly by brute force for small networks and small $k$-values.  For larger cases, the expected value is approximated by sampling a finite subset of these possibilities. The strong law of large numbers (SLT) guarantees this finite approximation converges to the actual value as the number of subsets indepentently samples from this distrubtion increase. One can then use the Central Limit Theorem (CTL), which states that the mean of $N$ sample values of $\avgdist(S)$ will tend towards a normal distribution, regardless if the sample distribution itself is normal or skewed. If $\mu$ and $\sigma$ denote the mean and standard deviation of the original samples, then the expected value and deviation of the sample means of is given by $\mu^* = \mu$ and $\sigma^* = \sigma/\sqrt{N}$, respectively.  This typically holds true for sample sizes $N \geq 30$, and  we generally consider $N \approx 100$ for our approximations.
We use the notation $\exptvalueapprox(k)$ to the denote this approximation:
\begin{equation}
\exptvalueapprox(k) \defined  \frac{1}{N} \: \sum_{i=1}^{N} \avgdist(S_i)   
\label{eqn:Ek}
\end{equation}
where $\{ |S_i| = k \}$ and each $S_i$ is random set of $k$ vertices.

\section{Experiments \& Methodology}
\label{sec:experiments}

For these experiments, we focused on connected simple graphs that were undirected, unweighted, with no self-loops or multi-edges. This represents the least common denominator for graph topologies, as not all datasets have edge weights and other metadata. Directed graphs were represented as undirected by making each edge bi-directional. For disconnected networks, we used the largest connected component. Additionally, the input networks had their vertices renumbered to be contiguous for optimized operations, and therefore did not necessarily match the vertex numbers in the original sources.

\begin{table*}[h]
\centering
    \caption{Application network topologies used in this study (largest connected component of undirected graph). The average degree is $\langle d \rangle$ and the maximum degree is $\Delta$.
     } 
  \vspace{0.1in}
    \begin{tabular}{ | l |  l r r r r r |}
    \hline 
    Network & Application  &  $|V|$  &   $|E|$   & $\avgdegree $ & $\Delta $ &  $\Delta/\avgdegree$   \\ [0.5ex]
    \hline \hline  
                         &   & &  & & &\\
   Zebra           & animal contact network & 23 & 105 & 9.13  & 14 & 1.5 \\
   Dolphin       &  animal contact network &62  & 159 & 5.13 & 12 & 2.3 \\
   Terrorist network & social network & 64 & 243 & 7.59 & 29 & 3.8\\
   High School & social network & 70 & 274 &  7.83 & 19 & 2.4\\
  MIT students & mobile social network &  96  & 2,539 & 52.90 & 92 & 1.7\\
Hypertext 2009 & social interaction & 113 & 2,196 & 38.9 & 98 & 2.5 \\
   Florida ecosystem wet        & food network  &  128 & 2,075 & 32.42 & 110 & 3.4 \\
   PDZBase &  metabolic network &  161 & 209  & 2.59 & 21 & 8.1 \\
  Jazz              &  collaboration network & 198 & 2,742  & 27.79 & 100 & 3.6\\
 GE\_200         & top-level web graph & 200 & 1,202 & 12,02 & 124 & 10.3\\
Chevron\_200 & top-level web graph & 200 & 5.450 &  54.50  & 189 & 3.5\\
Abilene218      & computer network & 218       & 226 & 2.07 & 10 &4.8\\  
Bethesda  & top-level web graph & 255 & 422 & 3.31 & 81 & 24.5 \\
\nn{C. Elegans}    & neural network &  297       & 2,148  &14.46 & 134 & 9.3 \\ 
NetScience & co-authorship & 379 & 914 & 4.82 & 34 & 7.0\\
Arenas-email & email communications & 1,133 & 5,451 & 9.62 & 71 & 7.4 \\
FAA
air traffic & infrastructure & 1,226  &  2,408  &  3.9  & 34 & 8.6\\
Human protein & protein interaction & 2,217 & 6,418 & 8.94 & 314 & 26.5 \\
ca-GrQc        & co-authorship &  4,158     & 13,422  & 6.46  & 81 & 12.5\\
ca-HepTh   &  co-autorship  &  8,638     &   24,806  &  5.74  &  65 & 11.3 \\ 
ca-HepPh & co-authorship & 11,204 & 117,619 & 23.38 & 491 & 21.0 \\
ca-CondMat & co-authorship & 21,363 & 91,286 & 8.54 & 279 & 32.6 \\
email-Enron & email communications & 33,696 & 180,811 &  10.73 & 1,383 & 128.9 \\
cit-HepPh & citation network & 34,401 & 420,784 & 24.46 & 846 & 34.6 \\
flickrEdges &             online social network             & 105,722 & 2,316,668 & 43.83 & 5,425 & 123.8 \\
email-EuAll & email communications & 224,832 & 339,925 & 3.02 & 7,636 & 2,525.3 \\ 
com-YouTube & online social network & 1,134,890  & 2,987,624 & 5.27 & 28,754 & 5,631.3 \\
soc-Pokec & online social network & 1,632,803 &  22,301,964 & 27.32 & 14,854 & 543.8 \\
soc-LiveJournal & online social network & 4,846,609 & 42,851,237 & 17.68 & 20,333 & 1,149.9 \\
[1 ex]
\hline
\end{tabular}
 \label{tbl:networks}
\end{table*}

The dataset comprised a wide range of application areas, including social, mobile, metabolic, neural, email, biological, and collaboration networks listed in Table~\ref{tbl:networks}. Examples were collected from network databases Konect~\cite{Konect}, SNAP~\cite{SNAP}, and UC Irvine~\cite{UCIrvine}, as well as several webgraphs generated by examining public websites. Network sizes ranged from less than 100 vertices (for exact verification of k-median problems) to networks with over 1 million vertices, with most networks  containing several thousand vertices. This study focused on 32 of these networks, comparing the eight algorithms from Section~\ref{sec:algos} for k-values from 1 to 100, resulting in roughly 25,000 experiments of graph, algorithm, and k-value combinations. This provided a clearer view of the performance landscape and algorithm behavior.

The evaluation occurred in three separate stages. First, we compared these heuristics to the exact solution for small networks and small k-values, where this could be computed by brute force. This established a baseline of how well these algorithms performed and with an actual error measurement. Second, for larger networks where a true solution was not feasible, we compared the methods against each other. Finally, we compared the methods to the null model of an expected value of A(S): the average solution obtained from a random guess (Eq.~\ref{eqn:Ek}).

Computational experiments were conducted on a  desktop workstation, running Ubuntu
Linux 5.15.0-46, with an AMD Ryzen 7 1700x (8-core) processor running at 3.4 GHz, and outfitted with 32 MB of RAM~\footnote{Certain commercial products or company names are identified here to describe our study adequately. Such identification does not imply recommendation or endorsement by the National Institute of Standards and Technology, nor does it imply that the products or names identified are necessarily the best available for the purpose.}. The algorithms were coded in C++, and compiled under GNU g++ 11.1.0 with the following optimization and
standardization flags: {\tt [-O3 -funroll-loops -march=native -std="c++11"]}. Modules were used
from the NGraph C++ library and Network Tookit~\cite{NGraph}, as well as optimized C++ implementations of
algorithms noted in the paper.

\section{Results}
\label{sec:results}

The results of the computational experiments on real networks showed significant variations. Despite claims made for any particular approach, we did not see a single method consistently producing a winner in every case.  Instead, we were presented with trade-offs for varying network  topologies. Nevertheless, we were able to form general observations about the expected behavior on classes of networks and provide some guidelines for choosing appropriate methods.

\subsection{Case study: distribution of $A(S)$}
\label{sec:distribution}

\begin{figure}
\begin{center} 
\includegraphics[width=\linewidth]{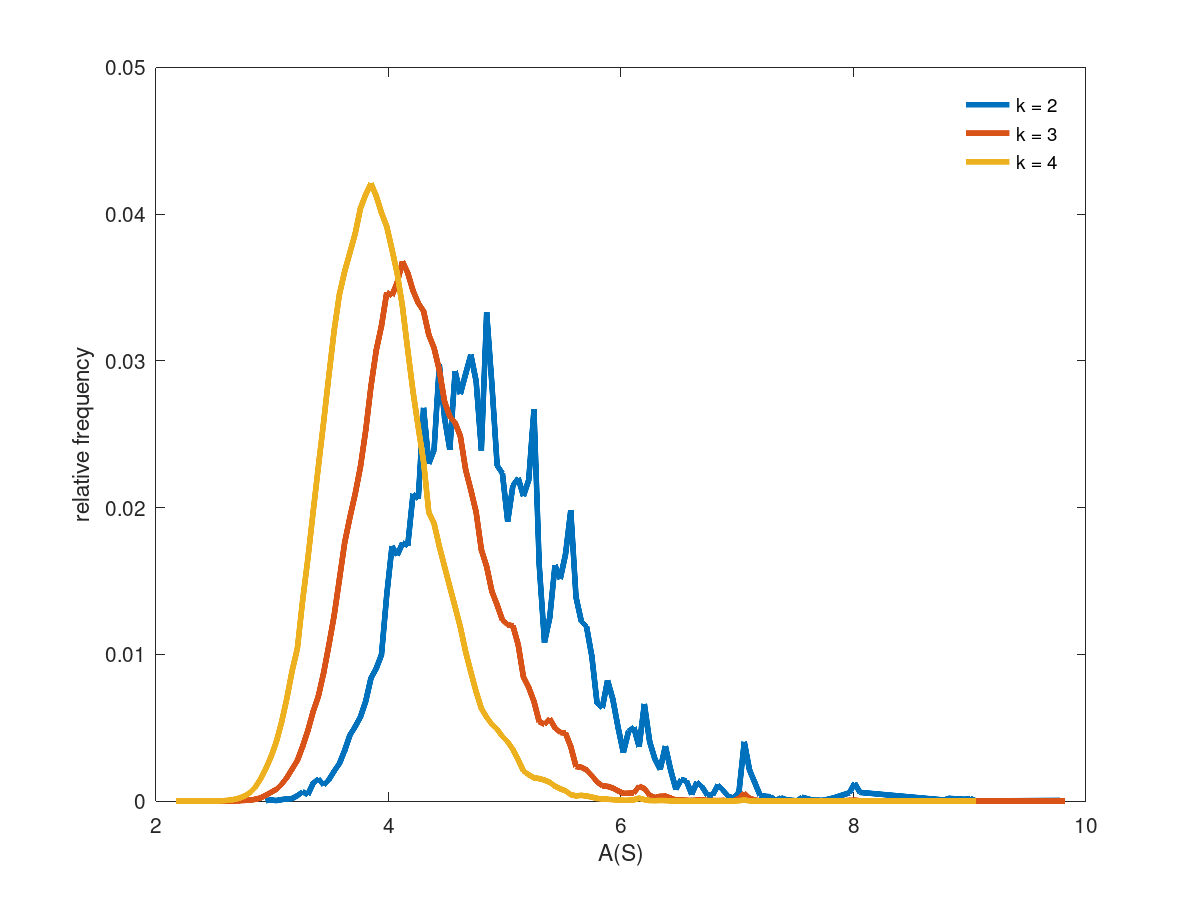} 

\end{center} 
\caption{Distribution of average distances $A(S)$ for all $k$-element subsets, $k=\{2, 3, 4\}$ in \netscience. 
Such example networks are used to determine the accuracy of approximations to  exact solutions.
} 
\label{fig:netscience:k234} 
\end{figure} 

To better understand the nature of the $k$-median problem, we can compute by $A(S)$ by brute force for {\bf all} of the $\binom{V}{k}$ $k-$element subsets.  It is a computationally expensive task, but insightful to appareciate the behavior of the function we are optimizing over.

As an example,  Figure.~\ref{fig:netscience:k234} illustrates the distribution of these values for of the {\em ca-netscience} network~\cite{SNAP}, using various values of $k$.  Here, $V=379$ and each curve signifies the distribution of $A(S)$ values for every one the possible $k$-element subsets. For instance, the $k=4$ curve represents the distribution of all $\binom{379}{4} = 846,153,126$ four-element subsets:
\begin{align*}
S_1 &= \{ 1, 2, 3, 4\} \\
S_2 &= \{ 1, 2, 3, 5\} \\
S_3 &= \{ 1, 2, 3, 6\} \\
    &\vdots \\
S_{846,153,126} &= \{ 376, 377, 378, 379\}
\end{align*}
In finding the \kmedian{} solution, we are interested in the left-most (minimum) $x$-value of the tail.  As $k$ increases, its distribution graph smooths out, as the number of samples points increase exponentially and begins to look more like a continuous distribution.  In this case, the optimal $4$-median, 
 $M(4) = 2.20$, with an expected value of $\exptvalue(4) = 3.97$. as given by Eq.~(\ref{eq:optimal}) and Eq.~(\ref{eq:expvalue}), respectively.

\begin{table}[t]
\centering
    \caption{\kmedian{} and expected values for \nn{ca-netscience}}
\begin{tabular}{rrrrr}
\\
\hline
$k$   &   $\optimalkmed(k)$   &   $\exptvalue(k)$ \ & \exptvalueapprox(k)  & $ \frac{\exptvalue(k)}{\optimalkmed(k)} $\\ 
\hline
\\
 1  &  3.90  & 6.04 & 6.01 & 1.55\\
 2 &  2.97  &   4.90 & 4.91 & 1.65\\
 3 & 2.53 & 4.33 & 4.31 & 1.71\\
 4 & 2.20  & 3.97 & 3.95& 1.80\\
 5 & 2.08   & 3.70 & 3.69 & 1.79
\end{tabular}
 \label{tbl:netscience:min:mean}
\end{table}

Table~\ref{tbl:netscience:min:mean} lists these values for $k = 1, 2, \ldots 5$.  Here, the {\bf average random guess can be less than twice the optimal \kmedian{} solution}, which is a much tighter result than for general problems.  (As described in later sections, this pattern is typical for complex networks.)  Furthermore, this table shows that the {\bf approximated expected value, $\exptvalueapprox(k)$ closely matches the exact value} within $\pm 0.5\%$ using the Central Limit Theorem with $N=100$.

\subsection{Comparisons with optimal \kmedian{} solutions}
\label{sec:opt:cmp}

\begin{figure*} 
\centering
\begin{subfigure}[t]{\scalefig \linewidth} 
    \includegraphics[width=\linewidth]{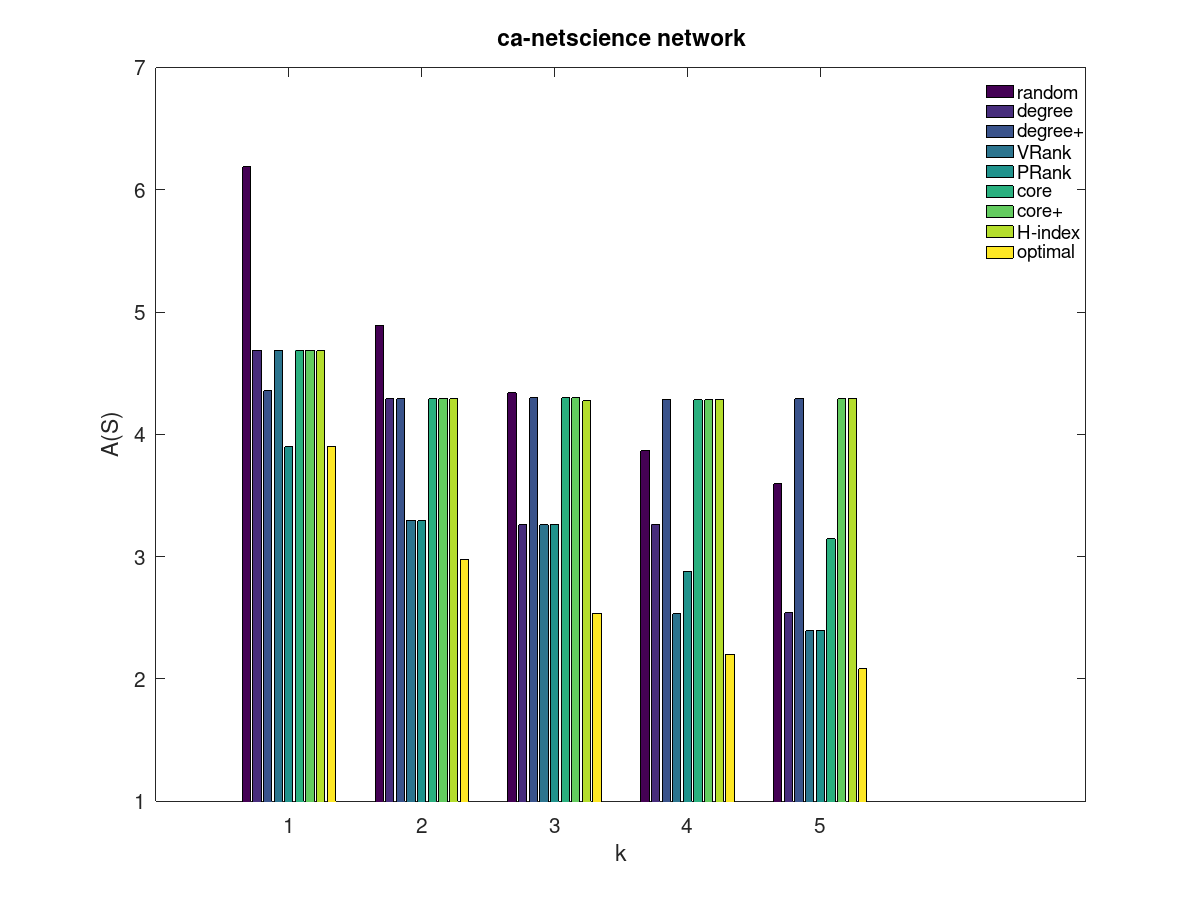} 
    \caption{\nn{ca-netscience} network:   \mvrank, and {\mprank{}} perform well but others are worse than random guessing.
    }
    \label{fig:netscience:opt} 
\end{subfigure}
\hfill
\begin{subfigure}[t]{\scalefig \linewidth} 
 	\includegraphics[width=\linewidth ]{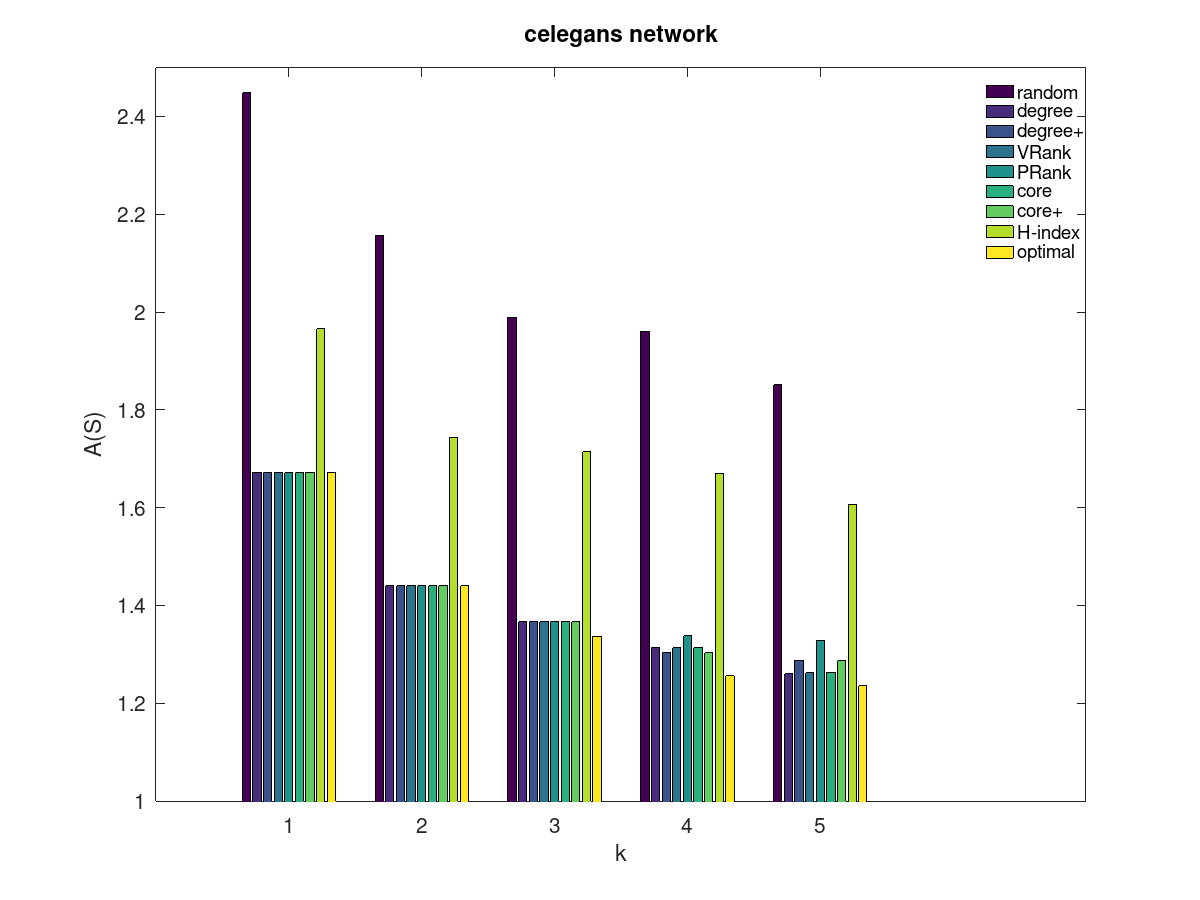} 
	\caption{\nn{C Elegans} network: methods work well but {\tt H-index} performs substantially worse.} 
	\label{fig:celegans:opt} 
\end{subfigure}
\hfill
\begin{subfigure}[t]{\scalefig \linewidth} 
 	\includegraphics[width=\linewidth ]{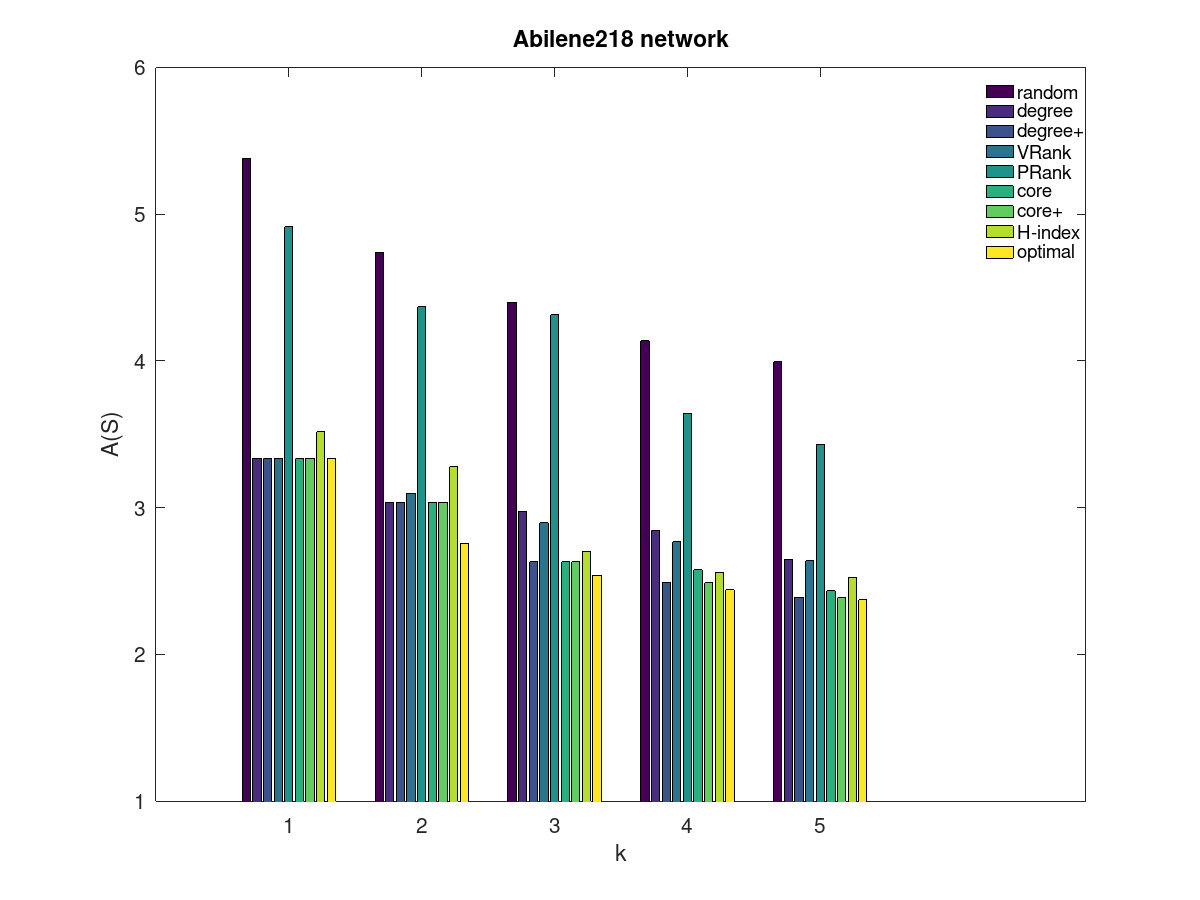} 
	\caption{\nn{Abilene218} network:  {\mprank{}} performs poor and not much better than random guessing.} 
	\label{fig:Abilene218:opt} 
\end{subfigure}
\hfill
\begin{subfigure}[t]{\scalefig \linewidth} 
 	\includegraphics[width=\linewidth ]{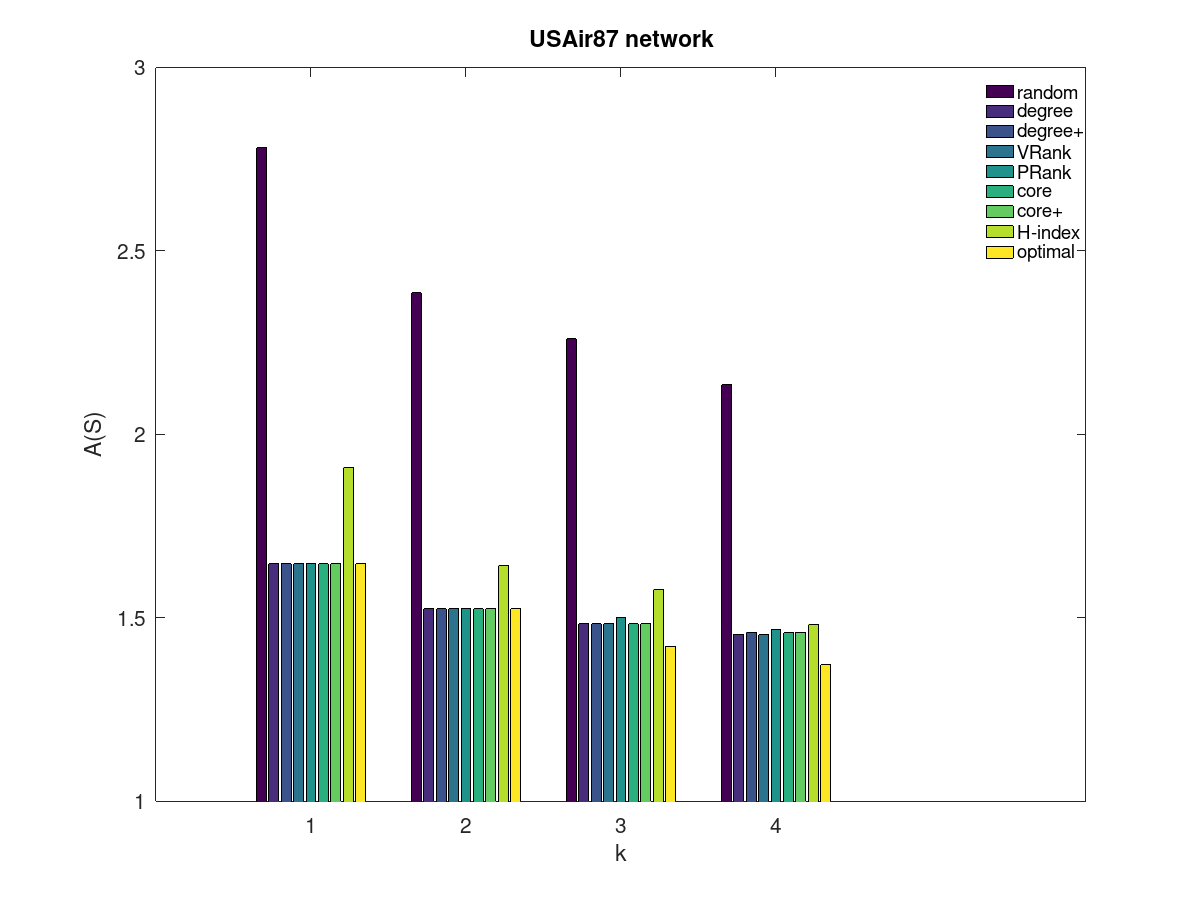} 
	\caption{\nn{USAir87} network:  approximation methods perform better than random guessing.} 
	\label{fig:USAir87:opt} 
\end{subfigure}
\caption{Comparison of approximation methods to exact \kmedian{} solutions for smaller networks.}
\label{fig:opt}
\end{figure*}

For small networks, we compared the accuracy of the approximation algorithms by running them for various values, up to $k = 5$, except where limited by the computation effort.
Figure~\ref{fig:opt}  illustrates the results for several small example networks.  The bar graphs show the computed approximations for $k = \{1, 2, \ldots, 5\}$ for the methods listed in Section~\ref{sec:algos}.  This collection illustrates several interesting patterns:
\begin{itemize}
\item   guessing a solution (\mrandom) performs, on average, within a factor of 2 from optimal
\item for $k > 2$ some approximation methods (\mdegreeplus, \mcore, \mcoreplus,  \mhindex) can perform {\bf worse} than random guessing
\item most methods (excluding \mrandom) stay within a 1.5 factor of optimal
\item \mvrank{} and \mprank{} seem to be perform best, staying within 1.2 of optimal for $1 \leq k \leq 5$.
\item \mcore, \mcoreplus, and \mhindex, typically perform worse, underperformed only by \mrandom
\end{itemize}
Although some networks exhibited slightly different behavior, these are the common patterns. We explore these characteristics in Sec.~\ref{sec:results:cmp}.

Compared to {\em ca-netscience},  the other networks appear better behaved. 
 The \nn{ C. Elegans} network (Fig.~\ref{fig:celegans:opt}) for example, shows extremely good approximations (relative error less than 5\%) for all methods, except for \mrandom{} and \mhindex{}.

In the \nn{Abilene218} network (Fig.~\ref{fig:Abilene218:opt}) we encounter quite different behavior: the \mprank{} method performs substantially worse than every other one, except random guessing.  This is in sharp contrast to other examples, where \mvrank{} and \mprank{} methods perform similarly and are often beat out other algorithms.

\begin{table*}
\centering
    \caption{Average error (\%) to true-optimal for small graphs ($1 \leq k \leq 5$)}
\begin{tabular}{lrrrrrrrr}
\hline
  Network         &  \small{\mrandom{}} &  \small{\mdegree{} }& \small{\mdegreeplus{}} & \small{\mvrank{}} & \small{\mprank{}} & \small{\mcore{}} & \small{\mcoreplus{}} &   \small{\mhindex{}} \\ \hline
Abilene218 & 68.9 & 11.1 & 3.4 &  10.2 & 54.0 & 4.4 & 3.4 & 8.4 \\ 
USAir87 & 60.0 & 2.6 & 2.7 & 2.6 & 3.1 & 2.7 & 2.7 & 10.7 \\ 
ca-HepTh & 46.4 & 4.7 & 5.8 & 4.7 &  4.7 & 26.2 & 26.2 & 34.9 \\ 
ca-netscience & 68.6 & 32.7 & 65.3 &  18.0 & 17.0 & 56.0 & 67.0 & 66.8 \\ 
celegans & 50.2 & 1.8 & 2.1 &  1.8 & 3.3 & 1.8 & 2.1 & 25.9 \\ 
faa & 48.4 & 10.7 & 5.0 & 10.7 &  8.9 & 4.8 & 5.3 & 12.6 \\ 
foodweb\_florida\_wet & 53.8 & 5.6 & 5.6 &  5.6 & 5.6 & 5.6 & 5.6 & 5.0 \\ 
hypertext\_2009 & 38.8 & 3.3 & 0.6 &  3.0 & 3.3 & 0.6 & 0.6 & 7.5 \\ 
jazz & 43.7 & 7.7 & 10.1 &  7.7 & 5.8 & 10.1 & 10.1 & 15.0 \\ 
pdzbase & 64.3 & 10.7 & 22.9 &  10.7 & 10.7 & 20.0 & 14.2 & 32.6 \\ 
\end{tabular}
 \label{tbl:small:opt}
\end{table*}

\begin{table}
\centering
    \caption{Ranking of methods by actual error (\%) in small graphs: average relative errors for each method in Table~\ref{tbl:small:opt}}
\begin{tabular}{lr}
\\
\hline
   method       &  error (\%) \\ \hline
   \\
\mvrank{} & 7.5 \\ 
\mdegree{} & 9.1 \\ 
\mprank{}& 11.6 \\ 
\mdegreeplus{} & 12.3 \\ 
\mcore{} & 13.2 \\ 
\mcoreplus{} & 13.7 \\ 
\mhindex{}  & 21.9 \\ 
\mrandom{} & 54.3 \\ 
\
\end{tabular}
 \label{tbl:optrank}
\end{table}

Finally, the {\em USAir87} network (Figure~\ref{fig:USAir87:opt}) illustrates that the approximation algorithms are capable of calculating good-quality solutions (even \mhindex{})  that are significantly better than random guessing.

Table~\ref{tbl:small:opt}  provides a tabular form of similar results from a larger study of 10 networks.
From this data we see that the behavior of these methods on real networks can vary significantly.   Ignoring \mrandom{} and \mhindex{} momentarily,  the  remaining competitive methods can be quite accurate for these networks. For example, {\em C Elegans}, {\em hypertext\_2009}, {\em USAir87} and {\em foodweb\_florida\_wet} all exhibit approximations that are within 5\% of optimal.The {
\em ca-netscience} network was the sole outlier, with the best methods of the group exhibiting roughly a 20\% error.

Taking the average error for each method across the networks  we arrive at (Table~\ref{tbl:optrank})  illustrating that \mvrank{} performed the best overall, with the other methods not too far behind. In this experiment, \mdegreeplus, \mcore, and \mcoreplus{} did not perform as badly, while \mhindex{} and \mrandom{} fared significantly worse.   This table represents the anaylis for small network comparisons with exact solutions.  It illustrates that the top methods generally work quite well,  typically {\bf within 10\% to 20\% of optimal} and seem  reasonable candidates for testing on larger networks.

\subsection{Analysis of larger networks}
\label{sec:results:cmp}
\begin{figure*}
\centering
\begin{subfigure}[t]{\scalefig \linewidth} 
    \includegraphics[width=\linewidth]{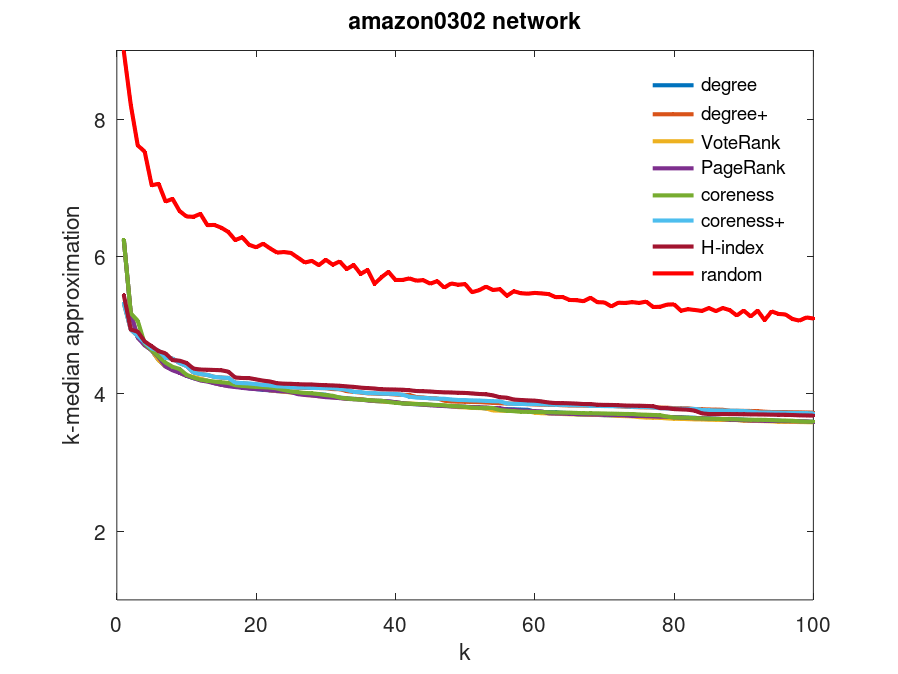} 
    \caption{{\em Amazon0302}: methods perform similar and an improvement over random guessing.}
    \label{fig:amazon0302} 
\end{subfigure}
\hfill
\begin{subfigure}[t]{\scalefig \linewidth} 
 	\includegraphics[width=\linewidth ]{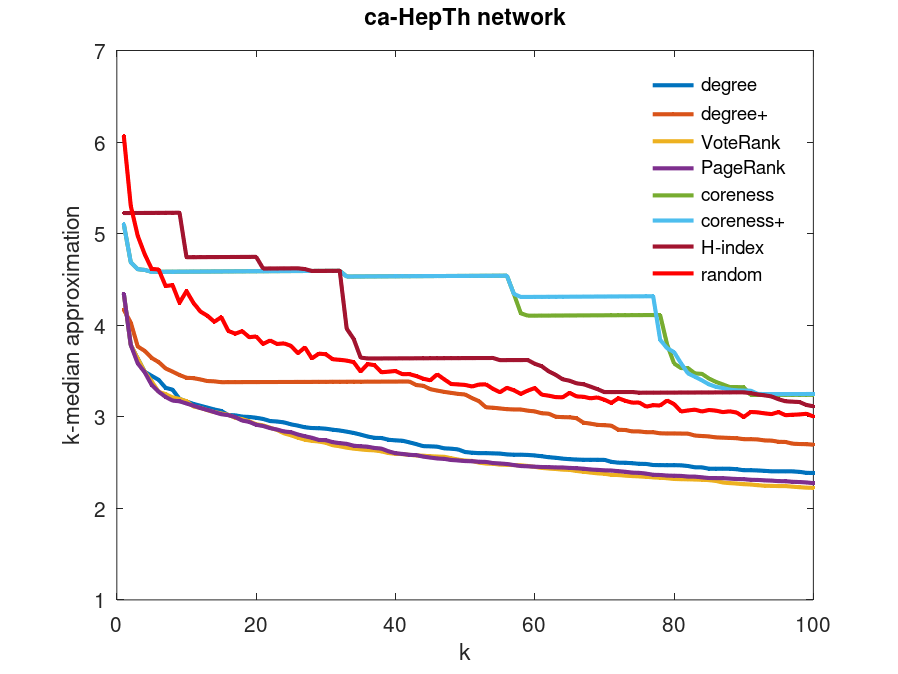} 
 	\caption{\nn{ca-HepTh}: \mcore{}, \mcoreplus{}, and \mhindex{} algorithms perform worse than \mrandom{}.} 
	\label{fig:ca-HepTh} 
	\end{subfigure}
\hfill
\begin{subfigure}[t]{\scalefig \linewidth} 
 	\includegraphics[width=\linewidth ]{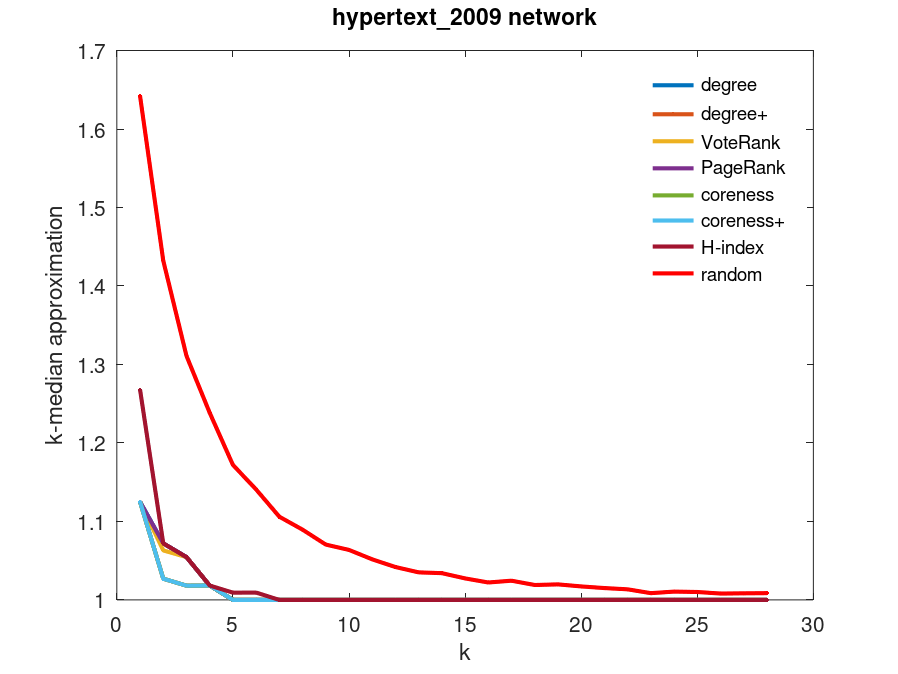}
 	\caption{\nn{ hypertex-2009}:  \mcoreplus{} closely outperforms most methods.} 
	\label{fig:hypertext-2009} 
\end{subfigure}
\hfill
\begin{subfigure}[t]{\scalefig \linewidth} 
 	\includegraphics[width=\linewidth ]{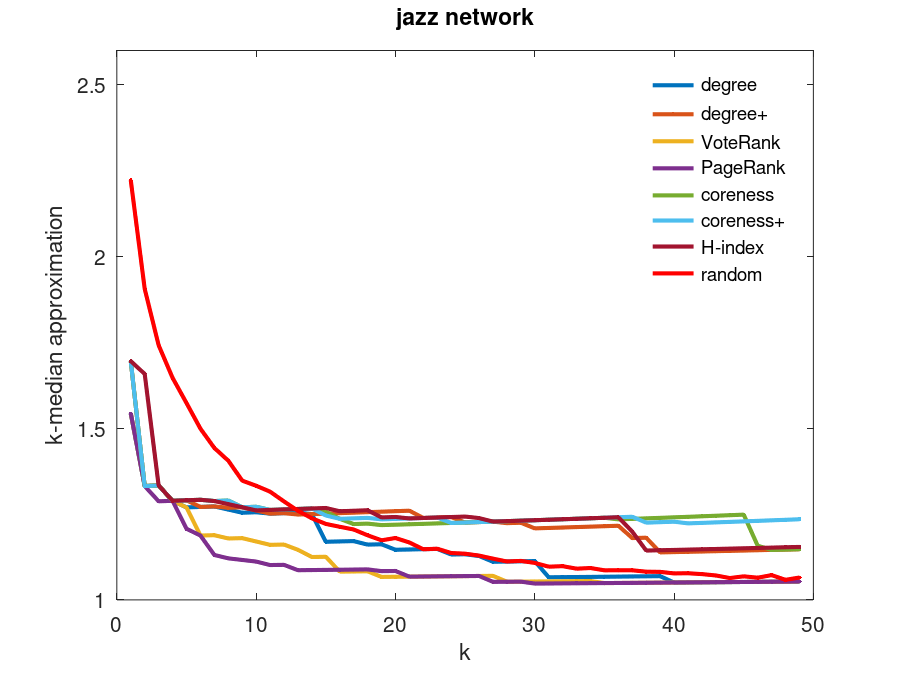} 
 	\caption{\nn{ jazz}: \mprank{} closely outperforms \mvrank{} and most other methods.} 
	\label{fig:jazz} 
\end{subfigure}
\hfill
\begin{subfigure}[t]{\scalefig \linewidth} 
	\includegraphics[width=\linewidth ]{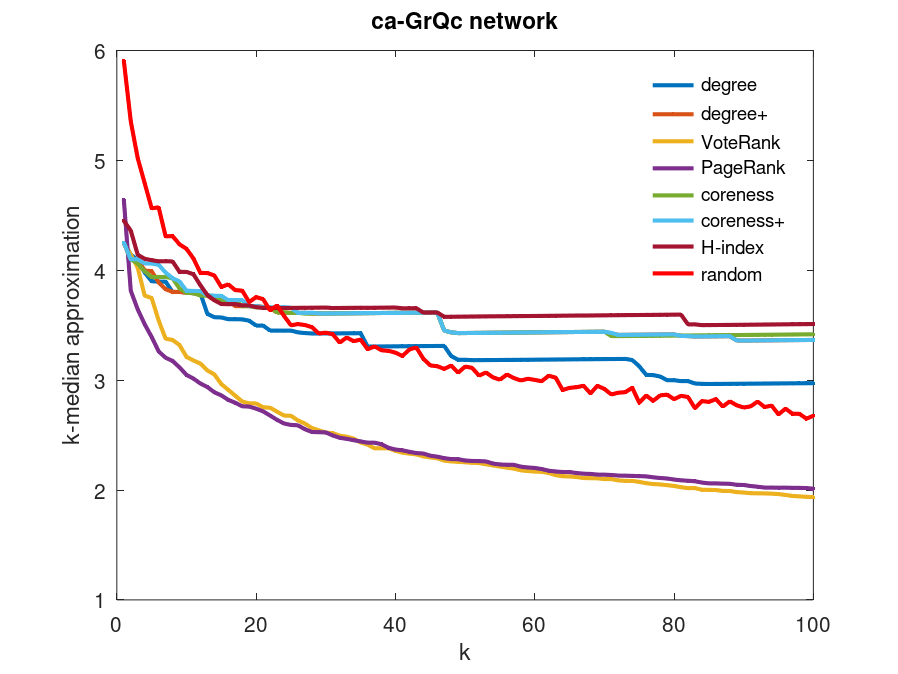} 
	\caption{\nn{ca-GrQc}: \mdegree{} falls behind \mrandom{} for $k > 25$.} 
	\label{fig:ca-GrQc} 
\end{subfigure}
\hfill
\begin{subfigure}[t]{\scalefig \linewidth} 
	\includegraphics[width=\linewidth ]{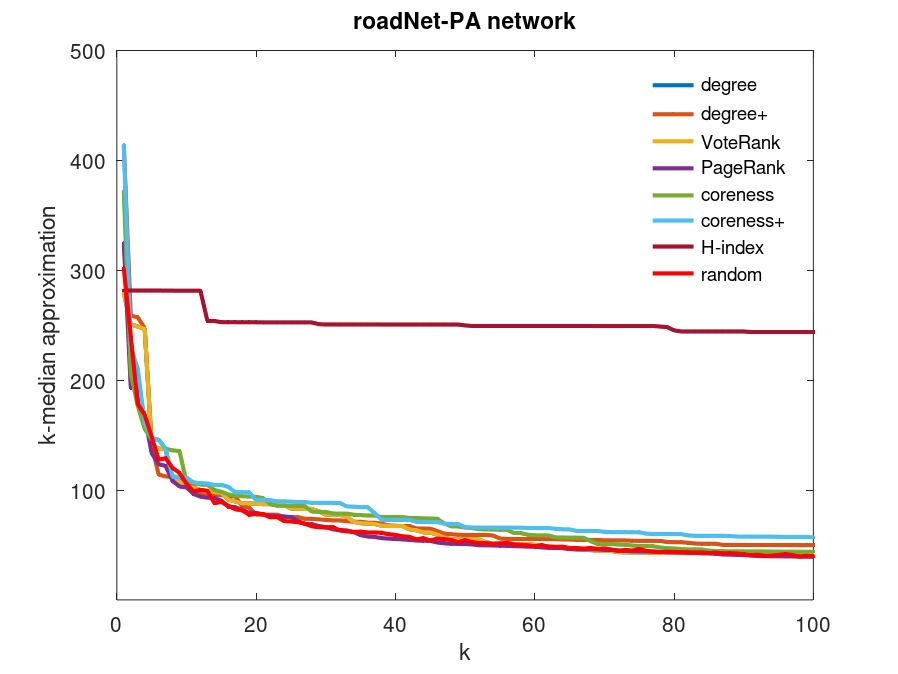} 
	\caption{ \nn{ roadNet-PA}: methods perform about as good as guessing and \mhindex{} is 5x worse.} 
	\label{fig:roadNet-PA} 
\end{subfigure}

\caption{Comparison of approximation methods to expected mean value of $A(S)$ for larger networks.}
\end{figure*}

This section describes the core of the study which focuses on larger networks more typical of modern datasets. 
In such cases,  the optimal solution is unknown and we use the expected value of a random guess to gauge the quality of the approximations.  Similarly to the smaller network examples, the expected-value \mrandom{}  provides solutions that are typically within a factor of two of optimal, so the bound can be useful in providing a practical measure.  As we shall see, a random guessing can sometimes {\bf outperform}  more sophisticated algorithms (although such occurrences are uncommon). 

In short, the outcome of the approximation algorithms on real network data yield mixed results, and examples can be found to substantiate nearly any hypothesis.  
As an illustration, we have listed several observations together with the specific network examples to substantiate that claim. In short, one can substantiate incomplete(and sometimes conflicting) hypotheses by simply choosing the appropriate network examples:
\begin{itemize}

	\item {\bf all approximation methods yield similar results}:  figure~\ref{fig:amazon0302} illustrates an example of the {\em amazon0302} network, where all approximation methods yield near identical results and there is little discernible quality difference between methods. Networks that exhibit this behavior include 
	\nn{as20000102},
	\nn{chevron-top200},
	\nn{com-YouTube},
	\nn{dlmf},
	\nn{email-Enron},
	\nn{p2p-Gnutalla31},
	\nn{soc-Epinions},
	\nn{soc-Slashdot0922},
	\nn{wiki-Vote}.

	\item {\bf \mrandom{} sometimes may out-perform heuristics}: Figure~\ref{fig:ca-HepTh} demonstrates the \nn{ca-HepTh} network where several algorithms (\mcore{}, \mcoreplus{}, \mhindex{}) perform substantially worse than expected value.  Other networks experienced similar behavior, where one or more  methods performed worse than random guessing:
\nn{ca-GrQc},
\nn{ca-HePh},
\nn{caj-GrQc},
\nn{ca-netscience},
\nn{faa}, and
\nn{pdzbase}
.

	\item{\bf \mcoreplus{} can may outperform other methods}: despite the relative poor performance of \mcoreplus{} there are (rare) instances where it is competitive with other methods and retains a slight advantage for small $k$ values.  One such example is the \nn{hypertext-2009} network, as shown in Fig.~\ref{fig:hypertext-2009}.  It should be noted that this network is rather well-connected, with an $M(k)$ values less than 1.7, even for small $k$.   The $k$-median approximation quickly goes to 1.0  (the minimum) for $k >5$.

	\item {\bf best methods sometimes dependent on $k$ values}:  network {\em email-EuAll} 
	demonstrates 
	an example where \mcore{} and \mdegreeplus{} algorithms perform best for small $k$, but are surpassed by \mvrank{} and \mprank{} for $k > 20$.

\item {\bf \mprank{} and \mvrank{} often perform equally well}: despite the criticism of \mprank{} for these types of problems, in practice it often performs just as good as \mvrank{}.  Example networks exhibiting this behavior include
\nn{as20000102},
\nn{bethesda},
\nn{ca-CondMat},
\nn{ca-GrQc},
\nn{ca-netscience},
\nn{faa}, and
\nn{human-protein}. 
Figure~\ref{fig:jazz} illustrates the \nn{jazz} network, where \mprank{} slightly outperforms \mvrank{}.

\item {\bf  \mdegree{} may performs worse than \mrandom{}}: although the \mdegree{} algorithm typically performs quite well, it can fall behind random guessing in rare instances.  The \nn{ca-GrQc} network (Fig.~\ref{fig:ca-GrQc}) exhibits such behavior.  Note that this network proved difficult for the \mhindex{}, \mcore{}, and \mcoreplus{} algorithms as well.

\item {\bf \mhindex{} typically worst performer}: although not specifically design for this purpose, \mhindex{} has become a popular centrality in the publishing industry and is often cited as a measure of importance for spreading information.  Unsurprisingly, it often comes in last (slightly better than random guessing) but sometimes it can completely miss the mark.  In the \nn{roadNet-PA} network, for example, it is roughly five times worse than \mrandom{} (Fig.~\ref{fig:roadNet-PA}).

\end{itemize}

\subsection{Case studies: million-node networks}
\label{sec:larger-networks}

\begin{figure}
\centering
\begin{subfigure}[t]{\scalefig \textwidth} 
    \includegraphics[width=\textwidth]{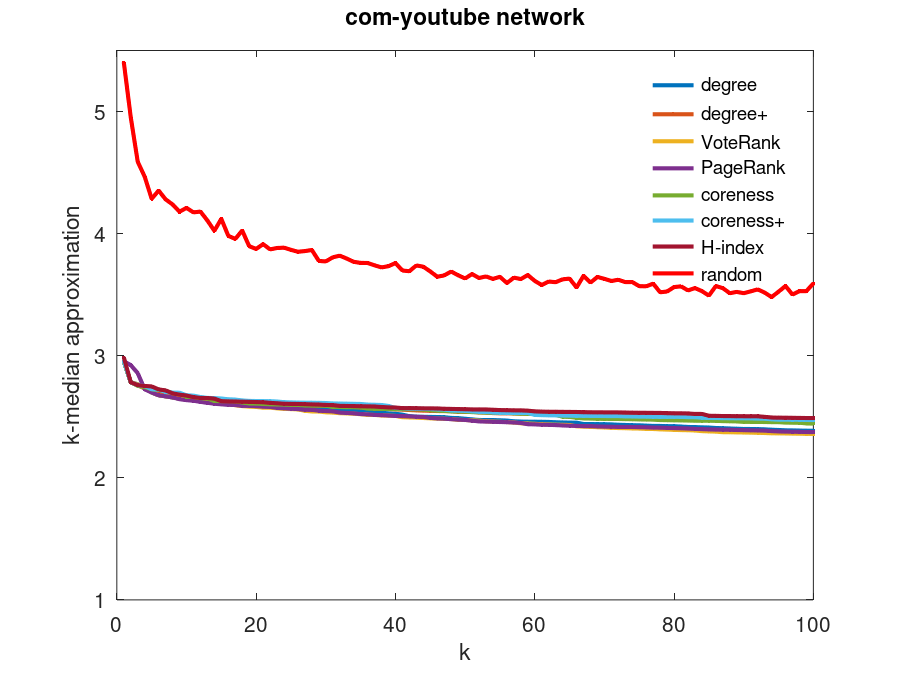} 
    \caption{ network {\em com-YouTube} $(V \approx 1.1M, \: E \approx 3.0M)$ 
}
    \label{fig:youtube} 
\end{subfigure}
\hfill
\begin{subfigure}[t]{\scalefig \textwidth} 
 	\includegraphics[width=\textwidth ]{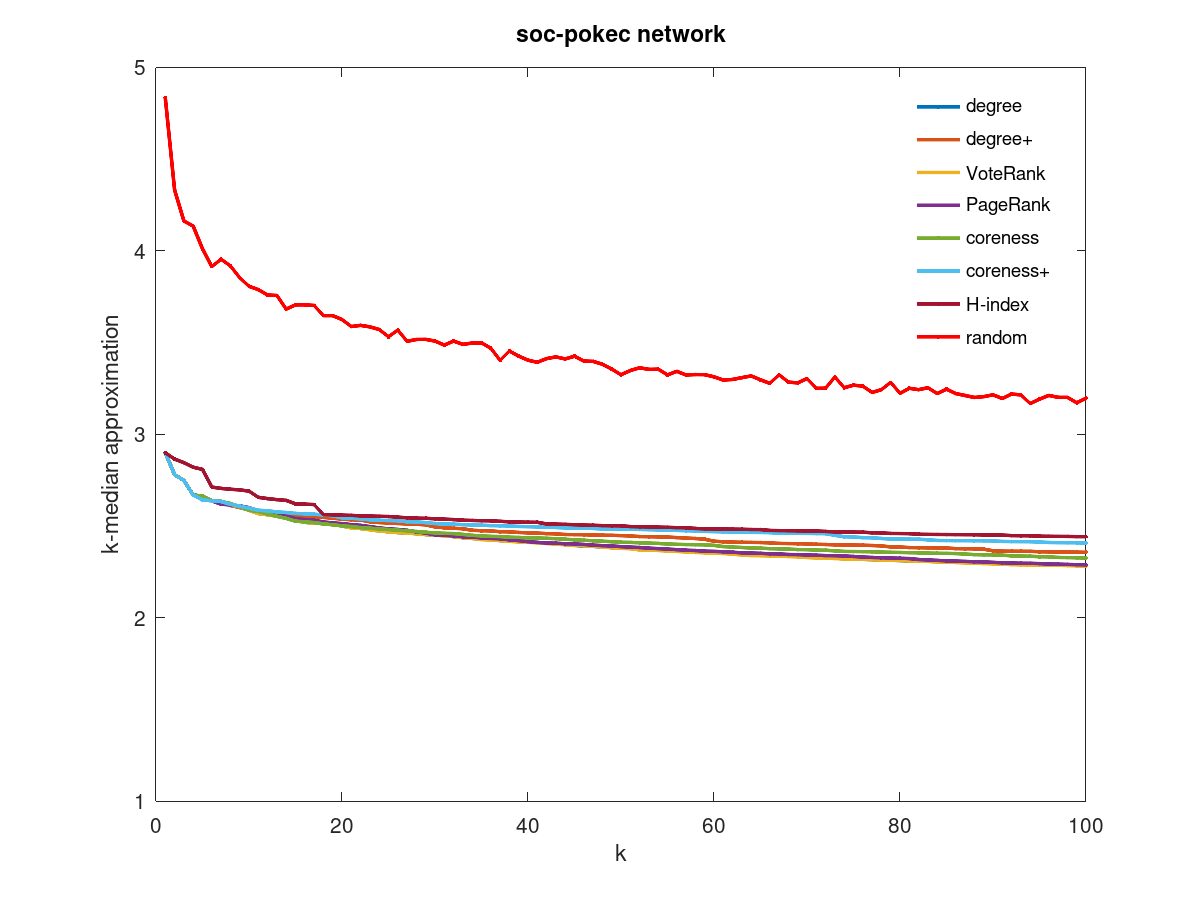} 
 	\caption{network {\em soc-pokec}  $(V \approx 1.6M, \: E\approx 22.3M)$  
} 
	\label{fig:soc-pokec} 
	\end{subfigure}
\hfill
\begin{subfigure}[t]{\scalefig \textwidth} 
 	\includegraphics[width=\textwidth ]{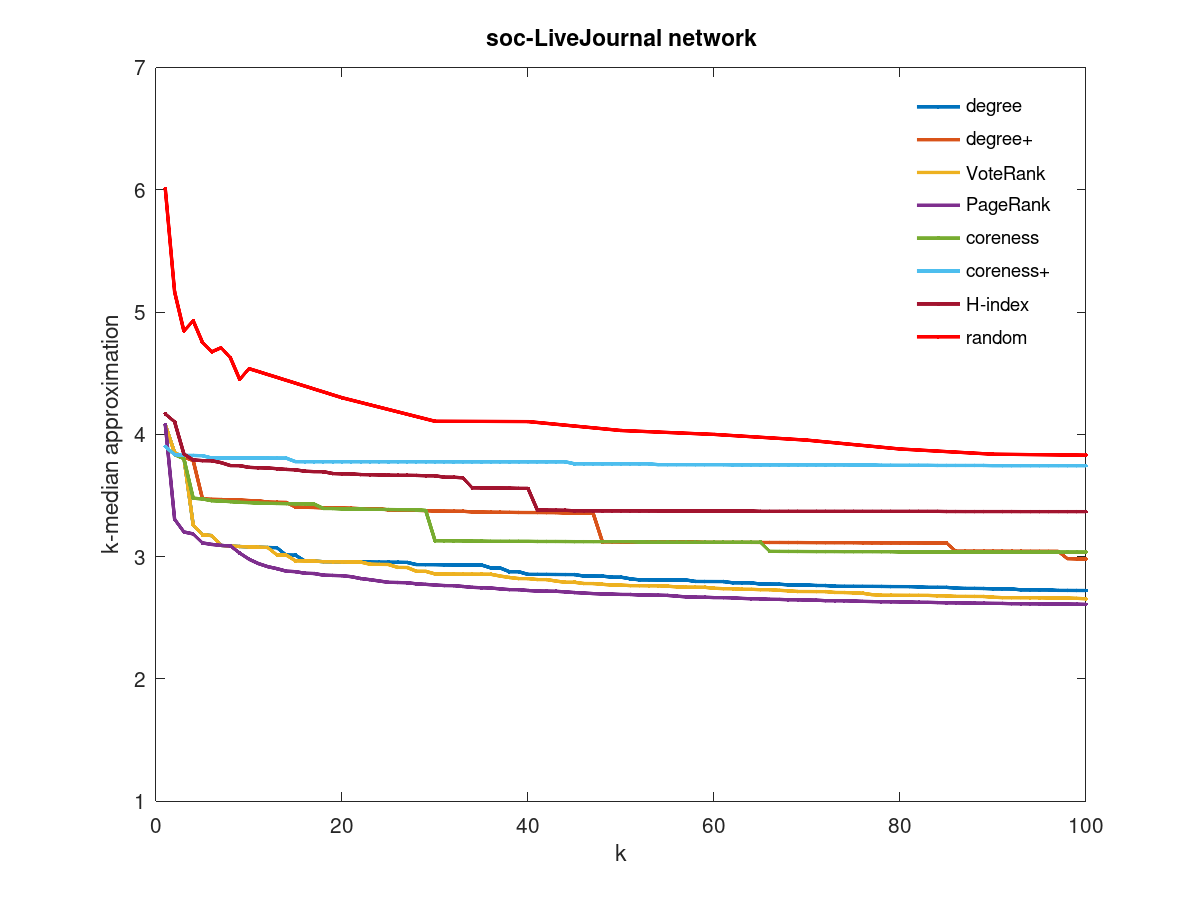}
 	\caption{network {\em LiveJournal}  $(V \approx 5M, \: E\approx 43M)$  where \mprank{}, \mvrank{}, and \mdegree{} algorithms perform significantly better
} 
	\label{fig:soc-LiveJournal} 
\end{subfigure}
\caption{Comparison of approximation methods for large online social networks.}
\end{figure}

Here we focus on three larger examples with millions of vertices and edges used in the study of large social networks~\cite{mislove-2007-socialnetworks}.
Figure~\ref{fig:youtube} illustrates the larger {\em YouTube} network (V=1,134,890 E=2,987,624) where the various heuristics perform about the same: roughly 35\% better than a random guess.  In this case, all methods yield nearly identical values, and one can simply use the fastest one (\mdegree) to generate competitive results.

Table~\ref{tbl:youtube} lists the how each method ranked in the top 1\%, 10\%, and 100\% of solutions.  For example, {\mdegree} scored in the top 10\% solutions about 3/4 of the time, while random guessing always remained within a factor 2 of the best solution.

Table~\ref{tbl:youtube:cost} illustrates the cost trade-off between computation time and solution quality.  Using the fastest method (\mdegree) as a reference, we see that \mvrank{} and \mprank{}  provided the best solutions, but at a computational cost of nearly three orders of magnitude.

\begin{table}
   \caption{Percentage of cases where each method scored within $x\%$ of best solution $(k = {1, \ldots ,100}$) for million-vertex network (YouTube).  Nearly all methods are within a factor of two of best solution.}
\begin{tabular}{lrrrr}
\\
\hline
   method         &  0\% (best) & 1\%   & 10\%  & 100\%  \\ \hline
\mdegree{} & 12.5 & 24.0 & 75.7 & 100.0 \\ 
\mdegreeplus{} & 8.5 & 11.5 & 42.7 & 99.9 \\ 
\mvrank{} & 20.4 & 44.2 & 91.9 & 100.0 \\ 
\mprank{} & 18.5 & 33.7 & 91.6 & 100.0 \\ 
\mcore{} & 9.1 & 15.5 & 50.1 & 99.0 \\ 
\mcoreplus{} & 6.3 & 8.7 & 41.0 & 97.8 \\ 
\mhindex{}  & 3.9 & 6.0 & 39.7 & 95.5 \\ 
\mrandom{} & 0.6 & 3.1 & 12.3 & 99.4 \\
\end{tabular}
 \label{tbl:youtube}
 
\end{table}

\begin{table}
\centering
    \caption{Relative computation cost for large graph (YouTube network) with over 1 million vertices and 3 million edges, together with overall performance within 10\% of best solution.}
\begin{tabular}{lrr}
\\
\hline
   method         &  cost factor & quality \\ \hline
\mdegree{} & 1.0 & 75.7 \\ 
\mdegreeplus{}  & 13.5 & 42.7\\ 
\mcore{} & 50.3  & 50.1\\ 
\mcoreplus{} & 112.8 & 41.0\\ 
\mhindex{}  & 259.1 & 39.7\\ 
\mprank{}  & 485.7 & 91.9 \\ 
\mvrank{} & 736.1  & 91.6 \\ 
\end{tabular}
 \label{tbl:youtube:cost}
\end{table}

Similar results are seen for the {\em soc-pokec} social network (V=1,632,803,  E=22,301,964) in Fig.~\ref{fig:soc-pokec}. From these two examples, one may be tempted to conclude that the algorithms perform equally well for large networks.
However, computations for the {\em LiveJournal} social network (V=4,846,609    E= 42,851,237) in Fig.~\ref{fig:soc-LiveJournal} show a significant difference between various methods, with \mprank{}, \mvrank{}, and \mdegree{} performing better than most other heuristics and roughly 30\% better than \mrandom{}.

\subsection{Overall results}

Table~\ref{tbl:best_large} describes the overall performance of approximation algorithms on the 32 networks under consideration.  The values are described as relative error to the best solution for each method.  For example, a value of 10 signifies that particular method performed on average within 10\% above the best possible heuristic  for each $k$ from one to one hundred. 

Table~\ref{tbl:rankings} presents the same results, based on their rankings, with the first place highlighted.  From here we see that \mvrank{} and \mprank{} come in first and second position, respectively, for the majority of cases while {\mdegree} comes in a close third position.

{\bf Table~\ref{tbl:overall:perf} summarizes these results}, where we compute the overall error of each method from the best solution for each $k$-value. For example,  \mdegree{} is typically about 5\% greater than the best solution, while \mrandom{}  produced, on average, a solution that was less than 50\% greater than the best algorithm.

\begin{table*}
\centering
    \caption{Quality of methods for large graphs ($k = 1,2, \ldots 100$).  Relative performance ( \%)  from best solution. A value of 10 signifies that on average that method performed  10\% above best possible value from all heuristics.}
\begin{tabular}{lrrrrrrrr}
\\
\hline
  Network         & \small{\mdegree{} }& \small{\mdegreeplus{}} & \small{\mvrank{}} & \small{\mprank{}} & \small{\mcore{}} & \small{\mcoreplus{}} &   \small{\mhindex{}}  & \small{\mrandom{}} \\ \hline
Abilene218 & 8.2 & 11.1 & 3.2 & 10.1 & 8.6 & 12.8 & 13.4 & 79.7 \\
USAir87 & 10.0 & 18.1 & 0.4 & 1.8 & 18.0 & 18.9 & 19.7 & 43.7 \\
amazon0302 & 0.5 & 3.0 & 0.3 & 0.4 & 0.8 & 3.0 & 3.9 & 48.0 \\
areans\_email & 3.3 & 9.4 & 0.0 & 0.8 & 7.3 & 11.1 & 10.9 & 30.9 \\
as20000102 & 1.4 & 12.5 & 0.0 & 0.5 & 4.5 & 11.4 & 7.0 & 76.1 \\
bethesda & 3.6 & 10.9 & 0.8 & 1.5 & 17.4 & 11.5 & 40.7 & 76.8 \\
ca-CondMat & 4.3 & 11.3 & 1.3 & 0.0 & 11.6 & 13.9 & 12.6 & 40.0 \\
ca-GrQc & 40.1 & 51.3 & 1.3 & 1.4 & 51.7 & 51.7 & 56.2 & 37.9 \\
ca-HepPh & 11.7 & 13.1 & 1.4 & 1.1 & 14.1 & 14.7 & 16.6 & 20.2 \\
ca-HepTh & 4.8 & 21.1 & 0.2 & 0.9 & 61.5 & 63.0 & 47.4 & 34.4 \\
ca-netscience & 12.6 & 34.9 & 0.3 & 3.2 & 35.6 & 59.5 & 48.0 & 50.9 \\
celegans & 0.9 & 6.9 & 0.1 & 0.9 & 1.6 & 6.6 & 7.8 & 25.0 \\
chevron\_top200 & 0.0 & 0.5 & 0.0 & 0.0 & 0.5 & 1.1 & 1.1 & 7.8 \\
cit-HepPh & 3.2 & 11.3 & 2.3 & 0.0 & 8.8 & 15.2 & 18.8 & 38.6 \\
com-youtube & 0.8 & 2.7 & 0.1 & 0.4 & 2.5 & 3.2 & 3.6 & 51.3 \\
dlmf & 3.1 & 7.6 & 0.0 & 2.0 & 8.5 & 7.5 & 8.3 & 67.2 \\
email-Enron & 2.7 & 11.6 & 0.7 & 0.4 & 9.0 & 12.1 & 13.3 & 59.1 \\
email-EuAll & 2.4 & 11.8 & 2.4 & 3.2 & 5.0 & 10.5 & 12.5 & 66.9 \\
faa & 8.6 & 31.0 & 0.7 & 1.8 & 18.0 & 41.1 & 30.8 & 40.3 \\
flickrEdges & 8.2 & 79.4 & 2.3 & 0.5 & 88.8 & 89.2 & 89.5 & 35.6 \\
foodweb\_florida\_wet & 0.1 & 0.6 & 0.1 & 0.1 & 0.4 & 0.7 & 1.5 & 8.7 \\
ge\_top200 & 3.0 & 14.3 & 0.3 & 0.3 & 9.4 & 17.4 & 22.8 & 36.0 \\
human\_protein\_gcc & 2.7 & 30.4 & 0.1 & 1.3 & 6.2 & 30.0 & 20.6 & 58.4 \\
hypertext\_2009 & 0.3 & 0.0 & 0.3 & 0.3 & 0.0 & 0.0 & 0.8 & 8.5 \\
jazz & 5.1 & 12.4 & 1.2 & 0.1 & 14.2 & 14.8 & 13.6 & 11.2 \\
p2p-Gnuetalla31 & 0.3 & 1.5 & 0.0 & 1.5 & 0.6 & 2.6 & 5.5 & 32.3 \\
pdzbase & 7.1 & 69.6 & 0.1 & 3.7 & 19.3 & 70.9 & 55.2 & 71.1 \\
roadNet-PA & 8.8 & 17.5 & 8.7 & 1.2 & 19.1 & 32.1 & 351.3 & 3.5 \\
soc-Epinions & 1.2 & 4.0 & 0.1 & 0.2 & 3.3 & 4.4 & 5.0 & 48.8 \\
soc-Slashdot0922 & 0.8 & 1.8 & 0.2 & 0.5 & 1.1 & 2.8 & 3.6 & 45.8 \\
web-Stanford & 2.2 & 21.3 & 0.8 & 0.7 & 12.4 & 20.9 & 26.5 & 37.0 \\
wiki-Vote & 1.6 & 2.0 & 0.6 & 0.1 & 2.8 & 2.0 & 2.2 & 38.5
\end{tabular}
 \label{tbl:best_large}
\end{table*}

\begin{table*}
\centering
    \caption{Ranking of k-median approximations}
\begin{tabular}{lrrrrrrrr}
\\
\hline
  Network         & \small{\mdegree{} }& \small{\mdegreeplus{}} & \small{\mvrank{}} & \small{\mprank{}} & \small{\mcore{}} & \small{\mcoreplus{}} &   \small{\mhindex{}}  & \small{\mrandom{}} \\ \hline
Abilene218 & 5 & 2 & \fbox{\bf 1} & 7 & 3 & 4 & 6 & 8 \\  
USAir87 & 3 & 4 & \fbox{\bf 1} & 2 & 5 & 6 & 7 & 8 \\  
amazon0302 & 3 & 5 & \fbox{\bf 1} & 2 & 4 & 6 & 7 & 8 \\  
areans\_email & 3 & 5 & \fbox{\bf 1} & 2 & 4 & 7 & 6 & 8 \\  
as20000102 & 3 & 7 & \fbox{\bf 1} & 2 & 4 & 6 & 5 & 8 \\  
bethesda & 3 & 4 & \fbox{\bf 1} & 2 & 6 & 5 & 7 & 8 \\  
ca-CondMat & 3 & 4 & 2 & \fbox{\bf 1} & 5 & 7 & 6 & 8 \\  
ca-GrQc & 3 & 4 & 2 & \fbox{\bf 1} & 5 & 6 & 8 & 7 \\  
ca-HepPh & 3 & 4 & 2 & \fbox{\bf 1} & 5 & 6 & 7 & 8 \\  
ca-HepTh & 3 & 4 & \fbox{\bf 1} & 2 & 7 & 8 & 6 & 5 \\  
ca-netscience & 3 & 5 & \fbox{\bf 1} & 2 & 4 & 8 & 6 & 7 \\  
celegans & 2 & 6 & \fbox{\bf 1} & 3 & 4 & 5 & 7 & 8 \\  
chevron\_top200 & \fbox{\bf 1} & 4 & 2 & 3 & 5 & 6 & 7 & 8 \\  
cit-HepPh & 3 & 5 & 2 & \fbox{\bf 1} & 4 & 6 & 7 & 8 \\  
com-youtube & 3 & 5 & \fbox{\bf 1} & 2 & 4 & 6 & 7 & 8 \\  
dlmf & 3 & 5 & \fbox{\bf 1} & 2 & 7 & 4 & 6 & 8 \\  
email-Enron & 3 & 5 & 2 & \fbox{\bf 1} & 4 & 6 & 7 & 8 \\  
email-EuAll & 5 & 3 & 6 & 7 & \fbox{\bf 1} & 2 & 4 & 8 \\  
faa & 3 & 5 & 2 & \fbox{\bf 1} & 4 & 7 & 6 & 8 \\  
flickrEdges & 3 & 5 & 2 & \fbox{\bf 1} & 6 & 7 & 8 & 4 \\  
foodweb\_florida\_wet & 3 & 5 & \fbox{\bf 1} & 2 & 4 & 6 & 7 & 8 \\  
ge\_top200 & 3 & 5 & \fbox{\bf 1} & 2 & 4 & 6 & 7 & 8 \\  
human\_protein\_gcc & 3 & 7 & \fbox{\bf 1} & 2 & 4 & 6 & 5 & 8 \\  
hypertext\_2009 & 5 & \fbox{\bf 1} & 4 & 6 & 2 & 3 & 7 & 8 \\  
jazz & 3 & 4 & 2 & \fbox{\bf 1} & 5 & 6 & 7 & 8 \\  
p2p-Gnuetalla3& 2 & 4 & \fbox{\bf 1} & 5 & 3 & 6 & 7 & 8 \\   
pdzbase & 3 & 7 & \fbox{\bf 1} & 2 & 4 & 6 & 5 & 8 \\  
roadNet-PA & 4 & 6 & 3 & \fbox{\bf 1} & 5 & 7 & 8 & 2 \\  
soc-Epinions & 3 & 5 & \fbox{\bf 1} & 2 & 4 & 6 & 7 & 8 \\  
soc-Slashdot0922 & 4 & 5 & \fbox{\bf 1} & 3 & 2 & 6 & 7 & 8 \\  
web-Stanford & 3 & 6 & 2 & \fbox{\bf 1} & 4 & 5 & 7 & 8 \\  
wiki-Vote & 3 & 4 & 2 & \fbox{\bf 1} & 7 & 5 & 6 & 8
\end{tabular}
 \label{tbl:rankings}
\end{table*}

\begin{table*}
\centering
    \caption{Efficiency of k-median approximations on large networks: computation time (secs)}
\begin{tabular}{lrrrrrrrr}
\\
\hline
   Network         & \small{\mdegree{} }& \small{\mdegreeplus{}} & \small{\mvrank{}} & \small{\mprank{}} & \small{\mcore{}} & \small{\mcoreplus{}} &   \small{\mhindex{}}  & \small{\mrandom{}} \\ \hline
\nn{soc-LiveJournal} & 0.01 & 0.8 & 50.8  & 35.2 & 5.4 & 11.5 & 5.6 & 166.9 \\
\nn{soc-pokec}& 0.01 &  0.5 & 24.4 & 20.0 & 2.2 & 4.7 & 3.28 &  62.5 \\
\nn{com-youtube} & 0.01 & 0.04 & 2.1 & 1.4 & 0.1 & 0.32 & 0.75 & 7.0
\end{tabular}
 \label{tbl:perf:sec}
\end{table*}

\begin{table}
\centering
    \caption{Ranked performance of k-median approximations. A value of  $x$ signifies that $M_{\mbox{\it method}}\:(k) $, on average, was within $x\%$ of the best solution for each graph and k-value combination from Table~\ref{tbl:best_large}.}
\begin{tabular}{lr}
\\
\hline
   method         & performance (\%) \\ \hline
\mvrank{} &0.9 \\ 
\mprank{}  &  1.3 \\ 
\mdegree{} & 5.1 \\ 
\mcore{} & 14.4 \\ 
\mdegreeplus{} & 16.7 \\ 
\mcoreplus{} & 20.5 \\ 
\mhindex{} & 30.3 \\ 
\mrandom{} & 41.6 \\ 

\end{tabular}
 \label{tbl:overall:perf}
\end{table}

\section{Conclusion}
\label{sec:conclusion}

  We have compared eight $k$-median approximation methods  for various $k$-values (typically 1 to 100) on 32 networks over a diverse range of application areas. After conducting thousands of experiments, we have observed patterns and formulated guidance for solving the k-median problem on a broad range of application network problems.

Overall, these approximation algorithms are efficient and some can produce good-quality solutions on complex networks. However, they cannot replace traditional methods\cite{allmethods} for general graphs that do not exhibit heavy-tailed degree distributions.  These skewed distributions often lead to {\em small-world} topologies, where the largest distance between vertices grows slowly compared to the size of the graph. In short, a vertex is typically not far from a large hub, quickly covering the graph and collapsing distances. In fact, the expected value of $E(k)$ for $k=1$  is  simply the average distance from any one vertex to the remainder of a given network topology, which is typically less than 6 in our experiments, thus providing another empirical demonstration of the ``six degrees of separation'' phenomena~\cite{watts1998collective}.

We have demonstrated that the algorithms in this study can indeed yield high-quality results on smaller networks where we can compute the optimal solution explicitly (Sec.~\ref{sec:opt:cmp}, Tbl.\ref{tbl:optrank}) with \mdegree{}, \mvrank{}, and \mprank{} achieving  roughly a  1.1 factor of the true solution.  By contrast, the best algorithms for general graphs provide a guaranteed factor of 3 or higher. 

For larger networks, the exact optimal solution is not computationally tractable and we can only compare the approximation methods against themselves. 
As Sec.~\ref{sec:results:cmp} illustrates, one  may reach an incomplete or  premature conclusions by examining only a small number of networks.  
Exploring a larger and more diverse dataset, however, reveals certain patterns that aid in algorithm choices.

Like many approximation heuristics, the practical question comes down to a trade-off between performance (computational cost) and quality of solution.  If one is willing to accept a factor of 2 from the best methods, then simply choosing $k$ random vertices from an uniform distribution may suffice. (This may come as a unexpected result, as hard problems typically do not behave in this manner.)  If a higher quality solution is needed, we can consult Tbl.~\ref{tbl:overall:perf} which summarizes the results of roughly 25,000 experiments.  Here we see that \mvrank{} and \mprank{} perform, on average, within about a 1.01 factor of the best method in every $k$-value in the [1:100] range.  The simple \mdegree{} method yields results on average within about 1.05 factor of the best method while exhibiting a performance speedup of three orders of magnitude over \mvrank{} and \mprank{}.

Thus, we can form a general best-practices guide for choosing the appropriate algorithms:
\begin{itemize}
\item if a quality factor of 2 is sufficient, choose $k$ random vertices from uniform distribution (\mrandom{}) 
\item if a better quality solution is needed, choose the top $k$ hubs (\mdegree{})
\item if quality still not sufficient, use \mvrank{} or \mprank{} for slight improvement (at a $10^2$ to $10^4x$ computational cost)
\end{itemize}
In practice, these approximation algorithms remain efficient, even for networks containing millions of elements.
The more expensive algorithms (\mvrank{},  \mprank{}) require about a minute   to approximate $k$-median solutions for up to $k =100$ on a personal computer (Tbl.~\ref{tbl:perf:sec}).  Thus, one possible approach would create an amalgamate {\em super-algorithm} which would run the seven methods (\mdegree{}, \mdegreeplus{}, \mvrank{}, \mprank{}, \mcore{}, \mcoreplus{}, \mhindex{}) concurrently and choose the best one for each $k$-value.   A final step could compare this to the expected value  (\mrandom{}) to give an indication how well the approximation methods  have improved the solution. 

In summary, the methods presented here do a reasonable job at estimating the $k$-median problem on complex networks. Despite the challenges of this fundamental problem, these methods provide a reasonable approximation and can be used efficiently to formulate approximations to this important problem, providing researchers with practical tools in studying large-scale complex networks.

\newpage

\newcommand{\bibname}{References}
{
	\footnotesize
\bibliographystyle{jresnist}
\bibliography{kmedian}
}

\end{document}